\documentclass[preprint,showpacs,preprintnumbers,amsmath,amssymb]{revtex4-1}
\usepackage{graphicx}
\usepackage{bm}
\usepackage{color}
\usepackage{amssymb}
\usepackage{epsfig}

\begin{document}
\bibliographystyle{}
.

\title{Eigenfunction statistics of Wishart Brownian ensembles}
\author{Pragya Shukla}
\affiliation{Department of Physics, Indian Institute of Technology,
Kharagpur, India}

\date{\today}


\begin{abstract}
 	
We  theoretically analyze the eigenfunction fluctuation measures for a Hermitian ensemble which appears as an intermediate state of the perturbation of a stationary ensemble by another stationary ensemble of Wishart (Laguerre) type. Similar to the perturbation by a Gaussian stationary ensemble, the measures undergo a diffusive dynamics in terms of the perturbation parameter but the energy-dependence of the fluctuations is different in the two cases. This may have important consequences for the eigenfunction dynamics as well as phase transition studies  in many areas of complexity where Brownian ensembles appear. 


\end{abstract}


\maketitle


\section{Introduction}

Eigenfunction correlations of linear operators play an important role in many  physical phenomenon e.g   phase transitions, transport, superconductivity, quantum entanglement, quantum chaos, atomic and nuclear reactions,  communication and networking etc (see e.g. \cite{gmw, been, mir, kota, me}). The complexity in a system leads to fluctuations of these correlations e.g. from one sample to another and it is necessary to consider their statistical behaviour. The  present study theoretically analyses the correlations of an important class of Hermitian operators i.e of type $A^{\dagger} A$,  with  operator $A$ representing a complex system  subjected to a single parametric random perturbation.  
The ensemble of these operators, known as Wishart ensembles \cite{wis} (also known as Laguerre ensembles),  have turned out to be successful models for  a wide range of areas e.g. multivariate statistical analysis \cite{and}, quantum chromodynamics \cite{vaar}, mesoscopic systems \cite{been, been1, zirn}, spin glasses \cite{chal}, financial systems \cite{stan, lal}, information theoretic studies \cite{ltbm} and communications \cite{com, mou, been, sm}, time-series analysis \cite{vp, vin2, seba}, biological networks \cite{lou}, geophysics \cite{abe} etc. Based on the nature of constraints on the matrix $A$ (originating from the exact symmetries and conservation laws in the  complex system it represents), Wishart ensembles (WE) can be of various types e.g stationary WE \cite{nagao}, correlated WE \cite{mou2, vin1, vp}, beta-WE \cite{satya}, non-white WE \cite{peche}, Brownian WE \cite{sp}; the present study concerns with last of these type i.e Brownian WE only.




 Originally introduced by Dyson to model the systems with partially broken symmetries and/or approximate conservation laws \cite{dy}, a Brownian ensemble (BE) is  one of the most simple basis-dependent ensembles  which appear as an intermediate state of crossover between two stationary ensembles i.e basis-invariant  random matrix ensembles \cite{dy, me, pijmp}.
%
Although Dyson's model was based on the assumption of Brownian dynamics of matrix elements due to thermal noise \cite{dy, me}, currently a Brownian ensemble is also  described as a non-stationary 
state of the matrix elements  due to a random perturbation of a stationary ensemble by another one. 
%
The type of a BE, appearing during the cross-over, depends on the nature of the stationary ensembles  and their different pairs may give rise to different BEs \cite{ap, fkpt, apps, sp, vp, psbe, pijmp}. 
(The present knowledge of ten types of universality classes \cite{zirn} of Hermitian matrix ensembles leads to possibility of many such cross-overs and, consequently, many types of BEs).  
Similar non-stationary states may also arise in other matrix spaces e.g. unitary matrix space e.g. due to a perturbation of a stationary circular ensemble by another one \cite{apps, vp}.

Brownian ensembles appear in many branches of physics \cite{gmw, been, mir, kota, me, fkpt, psand, ps-app, sp, vp, sups, pijmp} (see also references in \cite{vp, sp}). 
 In past there have been many studies of the Brownian ensembles (see for example \cite{me, rp, mp, apps, ap, sp, vp, fkpt, forres, mach, fgm, alt, ks, to, ls, chen1, shapiro, pich}, the list is by no means complete) but most of them are focused on the perturbation taken from a Gaussian stationary ensemble. (For the cross-overs, beginning from various stationary states e.g. GOE, 2GOE, Poisson, uniform etc and approaching GUE in infinite perturbation strength limit $\lambda \rightarrow \infty$, the $2^{nd}$ order correlation functions for all $\lambda$ have been explicitly evaluated \cite{ap}; for the other transitions the correlations are given implicitly by a hierarchic set of relations \cite{ap}). A detailed study of the  eigenvalue fluctuations for the Brownian ensembles in unitary matrix space, with stationary circular ensembles as the perturbation, was carried out in \cite{apps, vp}.   A similar analysis in Hermitian matrix space was discussed in \cite{sp}, extending the analysis of Gaussian type \cite{ap} to  Laguerre and Jacobi type. The study \cite{sp}  did not consider the eigenfunction fluctuations which however were analyzed in \cite{pswf, somm} in case of the Gaussian ensembles; (note the study \cite{pswf} is concerned with a multi-parametric Gaussian ensemble with BEs as a special case). The related information for other Hermitian types i.e Wishart and Jacobi is still missing.  This motivates us to pursue the present study   which is confined to  Wishart ensembles only due to technical reasons.
%

For last few decades, WEs have been subjected to extensive research investigations. 
Previous studies have primarily focused on their spectral statistical measures and many results for them  are  now known \cite{vin1, vp, nagao, mou2, peche, sm, bur, gao, mal, sil, agk, edel, fh}. But, notwithstanding a wide-ranging applicability,  the eigenfunction statistical measures of WE are theoretically known only in a limited number of cases e.g. basis-invariant cases in which statistics of the eigenvalues is uncorrelated with that of the eigenvectors. The reason for this information gap lies in the technical handicap: to derive the information about eigenvector statistics, it is necessary to integrate over eigenvalue-space. The correlations between eigenvalues and eigenvectors, usually present in the basis-dependent cases, make the integration mathematically complicated.  This handicap, in context of the basis-dependent Gaussian ensembles, was dealt with  by some approximations  which were later found to be in good agreement with numerical results \cite{pswf}.  This encourages us not only to apply the similar consideration in  present context but also attempt to extend their range of applicability. Our approach is based on  a diffusion equation  for the ensemble density i.e the distribution of Wishart-matrix elements in Hermitian matrix space. An essentially  similar approach for BE analysis was used by Kumar and Pandey in \cite{sp} however their interest being in  eigenvalue statistics only, they directly derived the diffusion equation for the eigenvalues, using 2nd order perturbation theory. As expected, an essentially same equation for the eigenvalues results from an exact diagonalization of the ensemble density diffusion equation. But the advantage of the latter is that  it leads to the diffusion equation for the eigenfunctions too. 
%



The paper is organized as follows. Section II presents a derivation of the 
diffusion equation for the elements of  Wishart matrix, say $L$, in Hermitian matrix space subjected to a random perturbation.   The equation gives the moments of the matrix elements which is used in section III to derive  the diffusion equations for the joint density of   eigenfunctions and eigenvalues of $L$. An integration of these equations over undesired variables then leads to evolution equations for various fluctuation measures of the eigenvalues and the eigenfunctions. The diffusion of eigenvalues is discussed in detail in \cite{sp};  a brief review  of main results for them in included in the appendix. Section IV  describes the derivation of the diffusion equations for the eigenfunction components which are used in section V to analyze the behaviour of their fluctuation measures.  Section V presents a summary of our main results.

\section{Brownian dynamics of matrix elements}

Consider an arbitrary $N_a \times N$  (with $N_a \ge N$) rectangular matrix $A_0$ subjected to a random perturbation, of strength $t$, by another $N_a \times N$  rectangular matrix $V$. The perturbed matrix $A(t)$ is described  as $A(t)=\sqrt{f} (A_0+t \; V)$ with $f=(1+ \gamma t^2)^{-1}$, $A(0)=A_0$ as a fixed random matrix and $\gamma$ as an arbitrary positive constant (\cite{me, ap}). 
Assuming the matrix elements of $A_0$ and $V$  distributed with the probability densities $\rho_0(A_0)$ and $\rho_v(V)$, the probability density $\rho_a(A) =\langle \delta\left(A-\sqrt{f} (A_0+t \; V) \right) \rangle $ of the $A$-ensemble is given by (with $\langle \rangle$ as the ensemble average)
\begin{eqnarray}
\rho_a(A) 
&=& \int \; \rho(A, t| A_0,0) \; \rho_0(A_0) \; {\rm D}A_0
\label{rhoh}
\end{eqnarray}
with  
\begin{eqnarray}
\rho(A, t|A_0,0) &=& \int \;  \delta\left(A-\sqrt{f} (A_0+t \; V) \right)\; \rho_v(V) \; {\rm D}V \nonumber \\
&=&
\left(\frac{1}{t \sqrt{f}}\right)^{N_a N} \; \rho_v \left(\frac{A-\sqrt{f} A_0}{t \sqrt{f}} \right);
\label{rho}
\end{eqnarray}  
here $A=A_0$ for $t \rightarrow 0$, $A \rightarrow V$ for $t \rightarrow \infty$. 

As discussed in \cite{sp}, the ensemble of matrices $A$ can lead to three important classes of Hermitian matrix ensembles: 
(i) Gaussian Brownian ensembles (GBE) of matrices $H =A+A^{\dagger}$ with $N=N_a$, 
(ii) Wishart (Laguerre) Brownian ensembles (WBE) with matrices $L=A^{\dagger} A$, and, 
(iii) Jacobi Brownian ensembles (JBE) of matrices $S$ which approach a form $S= (A^{\dagger} A+ B^{\dagger}B)^{-1/2} \; (B^{\dagger} B- A^{\dagger}A) \; (A^{\dagger} A+ B^{\dagger}B)^{-1/2}$.
Here $A$ and $A_0$ are real or complex for $L, H, S$ real-symmetric or complex Hermitian matrices, respectively.

 A wide-ranging applications of these ensembles make it desirable to study the statistical behavior of their eigenvalues and eigenfunctions. 
A recent study \cite{sp} describes the spectral fluctuations  for all three cases by a common mathematical formulation which is useful for a comparative analysis of their statistics, e.g the energy-dependence.  A similar approach  for the eigenfunction measures is however not available so far. Although the eigenfunction statistics for case (i) was analyzed in \cite{pswf}, an extension of those results to cases (ii) and (iii) is not directly obvious.  The present work analyses  the eigenfunction fluctuations  of  the Wishart case and describes the results in a common mathematical form applicable to Gaussian cases too. 




A variation of strength $t$ of the random perturbation $V$ leads to an evolution of the matrix elements $A_{kl}(t)=\sqrt{f} (A_{0;kl}+t \; V_{kl}(t))$ which, by a suitable choice of $\rho_v(V)$, can be confined to a finite space.  Here we consider $\rho_v(V)$ described by a Gaussian density: 
\begin{eqnarray}
\rho_v(V) =\left(\frac{1}{2 \pi v^2}\right)^{\beta N_a N/2} {\rm e}^{-{1\over 2 \; v^2} \; {\rm Tr} (V V^{\dagger})}
\label{rhov}
\end{eqnarray}
 where $V$ is real or complex for $L$ real-symmetric or complex Hermitian, respectively.  The first two moments of the matrix elements can then be written as 
\begin{eqnarray}
\langle{V_{kl;s}(t)} \rangle &=& 0, \qquad
\langle{ V_{kl;s}(t) \;  V_{mn;s'}(t')}\rangle
=  \beta \; v^2 \; \delta_{km} \; \delta_{ln} \; \delta_{ss'}  \; \delta(t-t')
\label{amn1}
\end{eqnarray}
with $\langle x \rangle $  implying an ensemble average of an arbitrary variable $x$.  The subscript "s" here refers to the number of components in a typical matrix element: $s=1 \rightarrow \beta$ with $\beta=1,2$ for $V_{kl}$  real or complex. 
For $V$-ensemble given by eq.(\ref{rhov}), the ensemble of matrices $V^{\dagger} V$ corresponds to the  stationary Wishart ensembles i.e Wishart orthogonal ensemble (WOE) for $\beta=1$ and Wishart unitary ensemble (WUE) for $\beta=2$. 

%


Our next step is to consider a diffusive dynamics of the matrix elements of $A$.  The  markovian character of the dynamics can be preserved if considered in terms of a rescaled parameter $Y=-{1\over 2 \gamma} \;  \ln f ={1\over 2 \gamma} \; \ln (1+ \gamma \; t^2)$ \cite{sp}:
\begin{eqnarray}
A(Y)  &\equiv&  A(0) \; {\rm e}^{-\gamma Y} +  V(Y) \; \left({1- {\rm e}^{-\gamma Y} \over \gamma}\right)^{1/2}
 \label{att0} 
\end{eqnarray} 
Using the above, a substitution of eq.(\ref{rhov}) in eq.(\ref{rho}) leads to  $\rho(A, Y| A_0,0)$ as a Gaussian. Alternatively, using the property that a convolution of two Gaussians is another Gaussian, one can write, for a small increment of perturbation strength at $Y$ with $\rho_v(V)$ given by eq.(\ref{rhov}),  
\begin{eqnarray}
A(Y+\delta Y)  &\equiv&  {A(Y) + \sqrt{2 \; \delta Y} \; V(Y) \over \sqrt{1 + 2 \; \gamma \; \delta Y} } \label{att} \\
&\approx &  A(Y) \; \left(1 -  \gamma \; \delta Y \right) +  \sqrt{2 \; \partial Y} \; V(Y) + O((\delta Y)^{3/2}). 
\label{atm1}
\end{eqnarray} 
Here the symbol $''\equiv''$ implies the equivalence of the ensembles of matrices on two sides.   
The ensemble  approaches to equilibrium as $Y \to \infty$. The equivalence of  eq.(\ref{att0}) and eq.(\ref{att}) along with the derivation of the diffusion equation for $A(Y)$ is 
discussed in \cite{sp}. 

%
%
%
%

As expected, the diffusive dynamics of the matrix elements $A_{kl}$ manifests itself in the $L$-matrix space and the moments for  the matrix elements $L_{mn}= \sum_{k=1}^{N_a} A_{km}^* A_{kn}$ can be calculated from those  of $A$. The above equations along with  relation between the elements of $L$ and $A$ gives the  moments of the matrix elements of $L$. As discussed in appendix B, the $1^{st}$ moment is same for both $\beta=1$ or $2$: 
\begin{eqnarray}
\langle{\delta L_{mn}} \rangle &=&  2 \; (\beta \; v^2 \; N_a \; \delta_{mn} -  \gamma \; L_{mn}  )\; \delta Y 
\label{lmn0}
\end{eqnarray}
but the $2^{nd}$ moment depends on $\beta$ as follows

\noindent{\bf Case $\beta=1$} 
\begin{eqnarray}
\langle\delta L_{mn} \; \delta L_{kl}^* \rangle  &=& \langle\delta L_{mn} \; \delta L_{kl} \rangle \nonumber \\
&=& 2 \; v^2 \; [L_{mk}  \; \delta_{nl} +L_{ml}  \; \delta_{nk} + L_{nk} \; \delta_{ml} + L_{nl} \; \delta_{mk}] \; \delta Y
\label{lmn1}
\end{eqnarray}

\noindent{\bf Case $\beta=2$} 
\begin{eqnarray}
\langle\delta L_{mn} \; \delta L_{kl}^* \rangle  
&=& 4 \; v^2 \; [L_{mk} \; \delta_{nl}  + L^*_{nl} \; \delta_{mk}] \; \delta Y \nonumber \\
\langle\delta L_{mn} \; \delta L_{kl} \rangle  
&=& 4 \; v^2 \; [L_{ml}  \; \delta_{nk}  + L^*_{nk} \; \delta_{ml}] \; \delta Y
\label{lmn2}
\end{eqnarray}
 %
with $\beta=1, 2$  for $L$ real-symmetric or complex Hermitian, respectively. 


Relevant information from the moments of $L$ can now be derived by using standard Fokker-Planck approach. In general, assuming Markovian process, the parametric diffusion of the  joint probability distribution $P_x(x_1,\ldots, x_N; Y)$ of $N$ variables $x_n$, $n =1,\ldots, N$ from an arbitrary initial condition, with $Y$ as the parameter, is given by 
\begin{eqnarray}
{\partial P_x\over\partial Y} \; \delta Y = {1\over 2} \sum_{k,l=1}^N {\partial^2 \over \partial x_{k} \partial x_l} \; (\langle\delta x_k \delta x_l \rangle\; P_x)  -\sum_{k=1}^N  {\partial \over \partial x_k} \; (\langle \delta x_k \rangle \; P_x)
\label{px}
\end{eqnarray}
Using the above approach, the diffusion of ensemble density $\rho_L(L)$ of the $L$ matrices from an arbitrary initial condition, say $\rho_{L_0}(L_0)$ with $L_0= A_0^{\dagger}. A_0$ can be described as 
%
 \begin{eqnarray}
{\partial \rho_L\over\partial Y}  = {\mathcal L} \rho + {\mathcal L}^* \rho 
\label{pxl}
\end{eqnarray}
where 
\begin{eqnarray}
{\mathcal L} \rho ={\beta^2\over 8} \sum_{k,l=1 \atop k \le l}^N  \sum_{m,n=1 \atop m \le n}^N  {\partial^2 ( B_2 \; \rho) \over \partial L_{kl}^* \; \partial L_{mn}} + {\beta^2\over 8} \sum_{k,l=1 \atop k \le l}^N  \sum_{m,n=1 \atop m \le n}^N  {\partial^2 ( B_1 \; \rho) \over \partial L_{kl} \; \partial L_{mn}} + {\beta \over 2} \sum_{m,n=1 \atop m \le n}^N  {\partial (B_ 0 \; \rho) \over \partial L_{mn}} 
\label{lxl}
\end{eqnarray}
where  $B_0=- 2 \; (\beta \; v^2 \; N_a \; \delta_{mn} -  \gamma \; L_{mn}  )$. But $B_1$ and $B_2$ depend on $\beta$: 
$B_1 = B_2 =  2\; v^2 \; [L_{mk} \delta_{nl} +L_{ml} \delta_{nk} + L_{nk} \delta_{nl} + L_{nl} \delta_{mk}] $ for $\beta=1$ and 
  $B_1=4 \; v^2 \; [L_{ml} \; \delta_{nk}  + L^*_{nk} \; \delta_{ml}]$, $B_2 =4 \; v^2 \; [L_{mk} \; \delta_{nl}  + L^*_{nk} \; \delta_{ml}]$, for $\beta=2$.
The stationary limit  ${\partial \rho_L \over  \partial Y} \to 0$ here corresponds to $Y\to \infty$ ((equivalent to $t \to \infty$ and $f \to 0$) and $L$ approaching $V^{\dagger} V$. As the  latter corresponds to WOE or WUE for $\beta=1, 2$ respectively, this is also the large $Y$-limit of $\rho_L(L)$.

The diffusion of the matrix elements of $L$ manifests itself in the dynamics  
 of its eigenvalues and eigenfunctions.
The  evolution equation for the  joint probability density  function (JPDF) of all eigenvalues and eigenvectors  can now be derived following the same steps as discussed in \cite{pswf} for a multi-parametric Gaussian ensembles. Although exact, this derivation is technically complicated  which  motivates us to present here an alternative route, physically motivating and technically easier: this is based on eqs.(\ref{lmn0}-\ref{lmn2})  to derive the  moments for the eigenvalues  and eigenfunctions which subsequently lead to their diffusion equations. 
%
Under Markovian dynamics assumption,  only the moments up to first order in $\delta Y$ are needed. The necessary steps are discussed in next section.


\section{first and second moments of eigenvalues and eigenfunctions}


 Let $U$ be  the $N\times N$ eigenvector matrix of $L(Y)$ , unitary in nature i.e $U^{\dagger} . U=1$ ($L$ being  Hermitian) and $\lambda$ be the $N\times N$ diagonal matrix of  its eigenvalues,  $\lambda_{mn}= \lambda_n \; \delta_{mn}$. A small change $\delta Y$ in parameter $Y$ changes $L$ and its eigenvalues  and eigenfunctions. Using standard 
perturbation theory for Hermitian operators and by considering matrix $L+\delta L$ in the eigenfunction representation of matrix $L$,  
a small change $\delta \lambda_n$ in the eigenvalue $\lambda_n$ can be given as 
\begin{eqnarray}
\delta \lambda_n = \delta L_{nn} +\sum_{m\not=n} {|\delta L_{mn}|^2 \over \lambda_n-\lambda_m}+
o((\delta X_{mn})^3)
\label{den}
\end{eqnarray}
where $L_{mn}=\lambda_n \; \delta_{mn}$ at value $Y$ of complexity parameter (due to $L+\delta L$ being considered in the diagonal representation of $L$). Eq.(\ref{den}) gives, up to first order of $\delta Y$ (see Appendix B),   
\begin{eqnarray}
\langle{\delta \lambda_n} \rangle
&=& 2 \; \beta \; v^2 \; \left[  N_a - {\gamma \over \beta \; v^2} \; \lambda_n +  \sum_{m=1,m\not=n}^{N} 
\; {(\lambda_n+\lambda_m)^{\nu} \over \lambda_n-\lambda_m}\right] \delta Y \nonumber \\
\langle{\delta \lambda_n \delta \lambda_m }\rangle &=& 
8 \; v^2 \;  \lambda_n \; \delta_{nm} \; \delta Y 
\label{enm}
\end{eqnarray}
where $\nu=1$; (the corresponding result for the Gaussian case is given by $\nu=0$ and is analogous to eq.(B4) and eq.(B5) of the appendix B of \cite{sp} with $Y=2\tau$). Note the moments given by eq.(\ref{enm}) are same as given by eq.(B.11) and eq.(B.12) in \cite{sp} with $\gamma=1$.


Our next step is to derive the moments of the perturbed eigenfunction components. 
%
The second order change in the $j^{\rm th}$ component $U_{jn}$ of an eigenfunction $U_{n}$ , in an arbitrary basis $|j \rangle$, $j=1,\ldots, N$,  due to a small change $\delta Y$ can  be described as 
  
\begin{eqnarray}
\delta U_{jn} = \sum_{m\not=n} {\delta L_{mn} \over \lambda_n-\lambda_m} U_{jm} +
\sum_{m,m'\not=n}^{N} { \delta L_{mn} \; \delta L_{m'n} 
\over (\lambda_n-\lambda_m)(\lambda_n-\lambda_{m'} )} U_{jm} \nonumber \\
-\sum_{m\not=n}^{N} {\delta L_{mn} \; \delta L_{nn}  
\over (\lambda_n-\lambda_m)^2} \; U_{jm} 
- {1\over 2} U_{jn} \sum_{m\not=n}^{N} {\delta L_{mn} \; \delta L_{nm}  
\over (\lambda_n-\lambda_m)^2} 
\label{f5}
\end{eqnarray}

To proceed further, we need to consider an ensemble average of eq.(\ref{f5}) which contains terms of type $\langle \delta L_{mn} U_{jm} \rangle$ and $\langle \delta L_{mn} \delta L_{kn} U_{jn} \rangle$. The calculation of these averages is easier if each $U_n$  is represented in a basis $|j \rangle$  in which $V(Y)$ is random, with $A(Y)$ and $V(Y)$ uncorrelated. The reason can be explained as follows:
as $U_n$ is an eigenvector of $L(Y) = A^{\dagger} (Y). A(Y)$ at $Y$ and $\delta L$ depends on the random perturbation $V(Y)$, latter assumed to be independent of $A(Y)$, the elements of $\delta L$ matrix are then   statistically independent of the components $U_{j n}$. (Also note the ensemble averaging  is over the ensemble of $\delta L$ matrices for a fixed $L$ at $Y$).  An appropriate choice for $|j \rangle$ for this purpose is the  eigenfunction basis of $L_0 \equiv A_0^{\dagger} A_0$ at $Y=Y_0$.   
%
%
Now using eqs.(\ref{lmn0}-\ref{lmn2}), it is easy to see that the ensemble averaged $U_{jn}$ has a non zero contribution only from the last term of eq.(\ref{f5}):

\begin{eqnarray}
\langle{\delta U_{jn}}\rangle &=&  - \beta \; v^2 \;    
\sum_{m=1,m\not=n}^N { (\lambda_n + \lambda_m)^{\nu} \over (\lambda_n-\lambda_m)^2} \; U_{jn} \; \delta Y 
\label{f6}
\end{eqnarray}
with  angular brackets implying conditional ensemble averages with fixed $e_j, U_j$, $j=1,\ldots, N$. 
%
%
But the $2^{nd}$ moment of the eigenvector components has a contribution 
only from the first term  in eq.(\ref{f5}) (up to first order in $\delta Y$)  and depends on $\beta$:

\noindent{\bf Case $\beta=1$} 
\begin{eqnarray}
\langle{\delta U_{jn} \; \delta U_{kl} } \rangle &=& 
2 \; v^2  \; \left( \sum_{m=1,m\not=n}^N {(\lambda_n+\lambda_m)^{\nu} \over (\lambda_n-\lambda_m)^2} \; U_{jm} \; U_{km} \; \delta_{nl} \;  \delta Y  
- { (\lambda_n+\lambda_l)^{\nu} \over (\lambda_n-\lambda_l)^2} \; U_{jl} \; U_{kn} \;
 (1-\delta_{nl}) \;\right)  \delta Y  
\nonumber \\
\label{f7q}
\end{eqnarray}

\noindent{\bf Case $\beta=2$} 
\begin{eqnarray}
\langle{\delta U_{jn} \; \delta U^*_{kl} } \rangle &=& 4 \; \; v^2  \; \sum_{m=1,m\not=n}^N {(\lambda_n+\lambda_m)^{\nu} \over (\lambda_n-\lambda_m)^2} \; U_{jm} \; U^*_{km} \; \delta_{nl} \;  \delta Y  
\nonumber \\
\langle{\delta U_{jn} \; \delta U_{kl} } \rangle &=& - 4\; v^2 \;  { (\lambda_n+\lambda_l)^{\nu} \over (\lambda_n-\lambda_l)^2} \; U_{jl} \; U_{kn} \; (1-\delta_{nl}) \;  \delta Y  
\label{f7p}
\end{eqnarray}

Further, to first order in $\delta Y$,  the ensemble averaged correlation between $\delta \lambda_k$ and $\delta U_{jn}$ is zero (for both $\beta=1$ or $2$): 
\begin{eqnarray}
\langle{\delta \lambda_k \; \delta U_{jn}}\rangle &=&  - 2 \; \beta \; v^2 \;  
\sum_{m=1,m\not=n}^N { L_{mn}\over (\lambda_n-\lambda_m)} \; U_{jn} \; \delta Y   = 0
\label{f7}
\end{eqnarray}
As mentioned above, the moments relations (\ref{f6}, \ref{f7p}, \ref{f7}) can also  be derived by an exact diagonalizaton of eq.(\ref{pxl}). 



Relevant information from the moments of eigenvalues and eigenfunction components of $L$ can now be derived by using standard Fokker-Planck approach (eq.(\ref{px}). As, for finite $Y$, the moments for the eigenfunction components depend on the eigenvalues too, we first write the diffusion equation for the joint probability density 
$P_{ef,ev} (\{U_n\}, \{\lambda_n \}; Y)$ at perturbation strength $Y$ where $\{U_n\}$ and $\{\lambda_n\}$ refer to the sets of all eigenvectors $U_1, \ldots, U_N$ and eigenvalues $\lambda_1, \lambda_2,..,\lambda_N$:

\begin{eqnarray}
{\partial P_{ef,ev} \over\partial Y} &=& ( L_U + L_U^*+ L_E) P_{ef,ev} 
\label{f8} 
\end{eqnarray}
where $L_U$ and $L_E$ refer to two parts of the Fokker-Planck operator 
corresponding to eigenvalues and eigenfunction components, respectively. Here $L_U$ 
is given as 

\begin{eqnarray}
L_U &=& {\beta \over 2}\sum_{j,n=1}^N {\partial \over \partial U_{jn}}
\left[{\beta\over 4} \sum_{k,l=1}^N \left({\partial \over \partial U_{kl}} \langle \delta U_{jn} \delta U_{kl} \rangle
+ {\partial \over \partial U^*_{kl}} \langle \delta U_{jn} \delta U^*_{kl} \rangle\right)
- \langle \delta U_{jn} \rangle \right] 
\label{f9}
\end{eqnarray}
and $L_E$ 
\begin{eqnarray}
L_E &=& \sum_{n} {\partial \over \partial \lambda_n}
\left[{1\over 2} \; {\partial \over \partial \lambda_n} \langle (\delta \lambda_n)^2 \rangle
- \langle \delta \lambda_n \rangle \right] 
\label{f9p}
\end{eqnarray}
Note here $P_{ef,ev}$ is subjected to following boundary condition for $\nu=1$: $P_{ef,ev} \to 0$ for $U_{jn} \to \pm \infty, \lambda_n \to [0, \infty )$ for $j,n=1 \to N$; this follows 
because the higher order moments of the ensemble density are assumed to be negligible.

As mentioned in section II, $\lim_{Y \to \infty} \rho_L(L, Y)$ corresponds to stationary Wishart ensembles. In the stationary limit  ${\partial P_{ef,ev} \over  \partial Y} \to 0$ of eq.(\ref{f8}), $P_{ef,ev}$ is therefore expected to approach corresponding JPDF of eigenvalues and eigenfunctions. As, in this limit, the eigenvalue statistics is independent of that of  eigenfunctions  \cite{me}, one can write $P_{ef, ev} = P_{ef}(U_1, \ldots, U_N) \; P_{ev}(\lambda_1, \ldots, \lambda_N)$ with $P_{ef}$ and $P_{ev}$ as the joint densities of the eigenfunctions and eigenvalues respectively. This in turn gives $L_E P_{ev}=0$ and $L_U P_{ef}=0$. It is easy to verify now that  $P_{ev} \propto \prod_{n=1}^N \lambda_n^{(N_a-N-1)/2} \prod_{j > k}^N |\lambda_j - \lambda_k| \; {\rm e}^{-{N_a\over 2v^2} \sum_{j=1}^N \lambda_j }$. Further as the eigenvectors are independent in this limit, subjected only to  unitary constraint ($L$ being Hermitian), one has $P_{ef} = \delta( U^{\dagger} U -1)$.



A substitution of the moments (eqs.(\ref{enm}, \ref{f6}, \ref{f7q}, \ref{f7p}, \ref{f7})) in eq.(\ref{f8}) 
followed by latter's integration  over all undesired variables  
will then lead to an evolution equation for the joint probability density 
of the desired combination  of  eigenfunctions and eigenvalues. 
 In next two section, we 
 discuss the JPDF related to various combinations of the eigenfunction components along with their eigenvalues. Note the JPDF of the eigenvalues only is discussed in detail in \cite{sp};  for completeness sake, the main results for them are summarized in the appendix.

\section{Joint distribution of eigenfunction components }

An integration of eq.(\ref{f8}) over eigenvalues leads to  the diffusion equations for the  joint probability 
distribution of the components of different eigenfunctions. The corresponding  equations for  the Gaussian case are derived and  discussed in detail in \cite{pswf}. Here we follow the same steps of the derivation  as for the Gaussian case and, for ease of comparison,  try to keep the same symbols as far as possible.  But the approximations used here are now applicable under more generic conditions and the results are presented in a form applicable to both Laguerre as well as Gaussian ensembles. To avoid repetition, here we give only the steps which are different  from those given in \cite{pswf}.

\subsection{ Joint distribution of a given components of all eigenfunctions and all eigenvalues}

We first consider the joint distribution 
of a given component of all eigenvectors along with  their eigenvalues. 
 Let $P_{1N}(Z,E,Y)$ be the probability, at a given $Y$, of finding the 
 $j^{th}$ component $U_{jn}$ of the eigenfunctions $U_n$ of $L$ between $z_{jn}$ and  
 $z_{jn}+{\rm d}z_{jn}$ and the eigenvalues $\lambda_n$ between $e_n$
 and $e_n+{\rm d}e_n$ for $n=1\rightarrow N$ (with $Z\equiv \{z_{j1}, z_{j2},.., z_{jN}\}, E\equiv \{e_1, e_2,..,e_N\}$).  It can be expressed as 
 \begin{eqnarray}
 P_{1N}(Z,E,Y) = \int f_j(Z,E,U,\lambda)\;  P_{ef,ev} \;  \left( \prod_{j=1}^N  {\rm d}\lambda_n \; {\rm D}^{\beta} U_n \right)
\label{pee}
 \end{eqnarray}
 with 
\begin{eqnarray}
f_j(Z,E,U,\lambda)=\prod_{n=1}^N 
 \delta(z_{jn} - u_{jn}) \delta^{\beta-1}(z^*_{jn} - u^*_{jn}) \delta(e_n-\lambda_n)
\label{fj}
\end{eqnarray}
and ${\rm D}^{\beta} U_n \equiv {\rm D} U_n$  for $\beta=1$ and 
${\rm D}^{\beta} U_n \equiv {\rm D} U_n {\rm D} U^*_n$ for $\beta=2$ where ${\rm D} U_n \equiv \prod_{k=1}^N \; U_{kn}$. 


As the $Y$-dependence in eq.(\ref{pee}) appears only through $P_{ef,ev}$, a differential of $P_{1N}$ with respect to $Y$ leads to 
an integral containing  the term $ {\partial P_{ef,ev} \over\partial Y}$. Substitution of  eq.(\ref{f8}), followed by repeated partially integration then leads to the  diffusion equation for $P_{1N}$. The intermediate steps are same as in the Gaussian case discussed in detail in \cite{pswf}. Proceeding along the same lines and using the limit $P_{ef,ev} \to 0$ at the end-points of the integration leads to the $Y$-governed diffusion equation of the joint probability density of the $j^{th}$  
component of all eigenvectors  and their eigenvalues,

 \begin{eqnarray}
 {\partial P_{1N} \over\partial Y}= \left( L_Z + L_Z^* \right) P_{1N} + L_E P_{1N}
\label{pp1}
 \end{eqnarray}
 where  $L_Z^*$ implies the complex conjugate of $L_Z$, with $L_Z=L_Z^*$ 
for $\beta=1$ case, and

\begin{eqnarray}
L_Z &=& {\beta \over 2}\sum_{n=1}^N {\partial \over \partial z_{jn}}
\left[{\beta\over 4} \sum_{m=1}^N \left({\partial \over \partial z_{jm}} \langle \delta z_{jn} \delta z_{jm} \rangle
+ {\partial \over \partial z^*_{jm}} \langle \delta z_{jn} \delta z^*_{jm} \rangle\right)
- \langle \delta z_{jn} \rangle \right] 
\label{pz1}
\end{eqnarray}
and $L_E$ given by eq.(\ref{f9p}). 
%
Substitution of correlations (\ref{f6}, \ref{f7p}) in eq.(\ref{pz1}) (with $z_{jn}$ replacing $U_{jn}$) and correlations (\ref{enm}) in eq.(\ref{f9p}) (with $e_j$ replacing $\lambda_j$) further leads to
\begin{eqnarray}
 L_Z &=& {\beta^2 \over 2^{\nu+2}} \sum_{n,m=1;n\not=m}^N 
 {(e_n+e_m)^{\nu}\over (e_n-e_m)^2} {\partial \over \partial z_{jn}}
 \left[ {\partial \over \partial z^*_{jn}} |z_{jm}|^2 
 -  {\partial \over \partial z_{jm}} z_{jn} z_{jm}  
 +  z_{jn}\right], \nonumber \\
 L_E &=& 
 \sum_{n} {\partial \over \partial e_n}\left[
  \beta \; a(e_n)  + \beta \sum_{m;m\not=n} {e_n^{\nu} \over e_m-e_n}
 + {\partial \over \partial e_n} \; e_n^{\nu}\right]. 
\label{aa1}
 \end{eqnarray}
 where  
\begin{eqnarray}
a(e)=\left({2\over \beta} \right)^{\nu}  \; e + \nu \; (N-1-N_a)/2.
\label{ae}
\end{eqnarray}
Note, for Gaussian case $\nu=0$, the above equation reduces to  (eq.(15) of \cite{pswf})).


\subsection{ Joint distribution of all components of an eigenfunction}

The probability distribution of the components $U_{nk}$,  of an eigenstate, say $U_k$ of  $L$  lying between $z_{nk}$ and $z_{nk}+{\rm d} z_{nk}$, with corresponding eigenvalue $\lambda_k$ between $e_k$ and $e_k+{\rm d}e_k$, $n=1 \rightarrow N$, can be given as  

 \begin{eqnarray}
 P_{N1}(z_{1k},...,z_{Nk}; e_k; Y) = \int\; \delta_k \; P_{ef,ev}\;  \prod_{j=1}^N  {\rm D}\lambda_j \; {\rm D}^{\beta} U_j, 
\label{pef}
 \end{eqnarray}
 where  
\begin{eqnarray}
\delta_k = \delta(Z_k - U_{k}) \delta^{\beta-1}(Z_k^* - U_{k}^*) \delta(e_k-\lambda_k).
\label{delk}
\end{eqnarray}
Partial differentiation of eq.(\ref{pef}) with respect to $Y$, subsequent substitution of  eq.(\ref{f8}) and repeated partially integration leads to the  diffusion equation for $P_{N1}$: 



\begin{eqnarray}
 {\partial P_{N1} \over\partial Y}= F_k+F_k^*+L_{e_k}P_{N1} 
\label{dpn1}
\end{eqnarray}
with 
\begin{eqnarray}
F_k &=& {\beta \over 2}\sum_{m=1}^N {\partial \over \partial z_{mk}}
\left[{\beta\over 4} \sum_{n=1}^N \left({\partial \over \partial z_{nk}} \; \langle \delta z_{mk} \delta z_{nk} \rangle
+ {\partial \over \partial z^*_{nk}} \langle \delta z_{mk} \; \delta z^*_{nk} \rangle\right)
- \langle \delta z_{mk} \rangle \right] 
\label{lfk}
\end{eqnarray}
and 
\begin{eqnarray}
L_{e_k} &=&  {\partial \over \partial e_k}
\left[{1\over 2} \; {\partial \over \partial e_k} \langle (\delta e_k)^2 \rangle
- \langle \delta e_k \rangle \right] 
\label{lfk1}
\end{eqnarray}
Again substitution of eq.(\ref{f6}, \ref{f7p}) in eq.(\ref{lfk}) and eq.(\ref{enm}) in eq.(\ref{lfk1}) (with $z_{nk}, z_{mk}, e_k$ replacing $U_{nk}, U_{mk}, \lambda_k$ respectively) further leads to

\begin{eqnarray}
F_k &=& {\beta^2\over 4} \; \left[\sum_{m=1}^N{\partial \over \partial z_{mk}} \left[  z_{mk} \; Q_{mm;k}^{02} \right]
+ \sum_{m,n=1}^N{\partial^2 \over \partial z_{mk} \partial z^*_{nk}} Q_{mn;k}^{12} \right], 
\label{fk}  \\
 L_{e_k}P_{N1} &=& {\partial \over \partial e_k}\left[
\beta \; a(e_k) P_{N1} + {\partial (e_k^{\nu} \; P_{N1}) \over \partial e_k} + \beta \; T_1(e_k)\right]
\label{lek}
\end{eqnarray}
with
\begin{eqnarray}
Q_{mn;k}^{r s} &=& \left({1\over 2}\right)^{\nu} \; \sum_{j; j \not=k} \int {(e_k+e_j)^{\nu} \over (e_k-e_j)^s} \; (z_{mj} z^*_{nj})^r \; {P_{N2}} \; {\rm d}e_j {\rm D}^{\beta} Z_j, 
\label{qab}
\end{eqnarray}
and
\begin{eqnarray}
T_1 (e_k) &=& e_k^{\nu} \; \sum_{j; j \not=k} \int { {P_{N2}} \over (e_j-e_k)}  \;{\rm d}e_j \; {\rm D}^{\beta} Z_j, 
\label{tk}
\end{eqnarray}
For later reference, note that $T_1(e_k) = e_k^{\nu} \; Q_{mn;k}^{01}(\nu=0)$.  
%
%
Also note that the expression for $Q_{mn; k}^{0s} $ is independent of the subscripts $m,n$ but latter are still retained in the notation so as to use a common mathematical expression for both $Q_{mn; k}^{0s} $ as well as $Q_{mn; k}^{1s} $. 
Here  $P_{N2}=P_{N2}(Z_k,Z_j,e_k,e_j)$ is the joint probability density of all the components of two  eigenvectors $Z_j\equiv \{z_{nj}\}$ and $Z_k\equiv\{z_{nk}\}$ 
($n=1\rightarrow N$) along with  their eigenvalues $e_j$ and $e_k$, respectively: 
\begin{eqnarray}
 P_{N2} = \int\;  { \delta_k}\;{ \delta_j}\; P_{ef,ev}(\lambda_1,..\lambda_N, U_1,..,U_N) \; \prod_{l=1}^N  {\rm d}\lambda_l \;  {\rm D}^{\beta} U_l
\label{pn2}
 \end{eqnarray}
where $\delta_k$ is defined in eq.(\ref{delk}).
%
Note
\begin{eqnarray}
P_{N1}(Z_k, e_k)= \int 
\; {P_{N2}} \;  {\rm d}e_j \; {\rm D}^{\beta} Z_j 
\label{pef4}
\end{eqnarray}

Eq.(\ref{dpn1}) is derived  from eq.(\ref{f8}) without  any approximation.  A similar  equation for $P_{N1}$ but with $\nu=0$ was derived in \cite{pswf} for the Gaussian Brownian ensembles (see eq.(18) of \cite{pswf}); note, for $\nu=0$, eq.(\ref{lek})  is analogous to eq.(19) of \cite{pswf} with symbol $Q_{mn;k}^{0 1} $ now replaced by  $T_1(e_k)$.  

As clear from eqs.(\ref{fk}, \ref{lek}, \ref{qab}, \ref{tk}), the right side of eq.(\ref{dpn1}) contains functions which are not explicitly written in terms of $P_{N1}$. In case of Gaussian ensembles in \cite{pswf}, eq.(\ref{qab}) was  approximated as $Q_{ab;k}^{r s} \approx \frac{(N-1)^{1-r}}{ \Delta_k^s} \; (\delta_{ab}-z_{ak}^* z_{bk} )^r \;  P_{N1}$  with $\Delta_k$ as the local mean level spacing at energy $e_k$ (see eq.(22) of \cite{pswf}). The approximation was however based on an assumed  weak statistical correlation between the eigenvalues and the eigenfunctions. Here we consider its improvement to include more generic regimes, based on the following ideas: 
(i) the eigenvalues at a distance more than few mean level spacing are uncorrelated, (ii) the average correlation between components of an eigenfunction is almost same as another eigenfunction if their eigenvalues are approximately equal. Using these ideas, it can be shown that (see Appendix D for details)
\begin{eqnarray}
Q_{mn;k}^{r s} &\approx &   {\mathcal K}_s \; \; \left(\overline{\langle z_{nk} z_{mk}^*\rangle} \right)^r  \; P_{N1}(Z_k, e_k)
\label{qaba2}
\end{eqnarray}
where 
\begin{eqnarray}
{\mathcal K}_s(e_k) &=&   \left({2\over E_c}\right)^s \; N \;  e_k^{\nu}, 
\label{kk}
\end{eqnarray}
where  $E_c$ is an important system-specific  spectral-range defined as follows: the eigenvalues at distances more than  $E_c$ around $e$,  are uncorrelated. (Note, in context of disordered systems, $E_c$ is also referred as the Thouless energy  and is of the order of local mean level spacing $\Delta_e$).  

Further, as mentioned below eq.(\ref{tk}), the expression for $T_1(e_k)$  is related to  $Q_{mn;k}^{01}(\nu=0) $, latter given by eq.(\ref{qab}). Thus $T_1(e_k)$ can be approximated as 
\begin{eqnarray}
T_1 (e_k) &\approx &  {\mathcal K}_1 \;P_{N1}(Z_k, e_k) 
\label{tk1}
\end{eqnarray}

Substitution of the approximations (\ref{qaba2})  in eqs.(\ref{fk}, \ref{lek}) helps to express $F_k$ in terms of  $P_{N1}$:
\begin{eqnarray}
F_k &=& {\beta^2 \; {\mathcal K}_2 \over 4 }  \; \sum_{n=1}^N   { \partial \over \partial z_{nk}}
\left[ \sum_m \overline{\langle z_{nk} z_{mk}^*\rangle}  \;\; {\partial P_{N1}  \over \partial z^*_{mk}} +   {z_{nk}} \; P_{N1}  \right]  
\label{fk2}
 \end{eqnarray}
With help of  eq.(\ref{fk2}), eq.(\ref{dpn1}) reduces now to a differential equation for $P_{N1}$ only. Here it must be noted that, for $\nu=0$, eq.(\ref{fk2}) gives $F_k$ for the Gaussian case. (The latter case was discussed in \cite{pswf} and eq.(\ref{fk})  approximately reduce to  eq.(23) of \cite{pswf} if one substitutes $\langle z_{nk} z_{mk}^*\rangle \approx {1\over N} \; \delta_{mn}$ in eq.(\ref{fk2}). The latter approximation is valid for almost extended eigenfunctions which was the basis of  derivation 
in \cite{pswf} ).



{\bf Stationarity limit}: as mentioned in section II, the matrix $L$ in this limit approaches $V^{\dagger} V$ with its statistics given by a stationary Wishart ensemble WOE or WUE. The distribution for the components of a typical  eigenfunction for these cases is known to be Gaussian. It is easy to verify that the solution of eq.(\ref{dpn1}) (along with eqs.(\ref{fk},\ref{lek}) in $Y \to \infty$ limit or, equivalently ${\partial P_{N1} \over \partial Y}=0$, is indeed a Gaussian. The steps are as follows:        
in this limit,  the distribution of eigenvalues and eigenfunctions are independent of each other and  one can write
$ Q_{mn;k}^{r2} = ({1\over 2})^{\nu} \; \sum_{j; \not=k} \langle (z_{mj}^* z_{nj} )^r\rangle \; \langle {(\lambda_j + \lambda_k) \over (\lambda_j -\lambda_k)^2} \; \delta(e_k-\lambda_k) \rangle$.
Further  $\sum_{j; \not=k} \langle (\lambda_j -\lambda_k)^{-2}  =c_0 \; (N-1)$  for a classical ensemble  (with $c_0$ a constant, see eq.(9.3.9) of \cite{me})
and the correlation  $\langle z_{mj}^* z_{nj} \rangle$ is independent of $j$ as well as the eigenvalue statistics:
$\langle z_{mj}^* z_{nj} \rangle = {1 \over N} \; \delta_{mn}$. Using these relations in eq.(\ref{qab}), one can write
$ Q_{mn;k}^{r2}  \approx c_0 \; e_k^{\nu} \; P_{N1}$. An $e_k$-integration of eq.(\ref{dpn1})  then leads to an equation with its solution  $\int P_{N1} \; {\rm d}e_k$ as a product of independent Gaussian distribution of the components $z_{nk}$, $n=1 \to N$. 



\subsection{ Joint distribution of the components of many eigenfunctions}

The joint probability density 
 $P_{Nq}$ of the components $U_{nk}$ ($n=1 \rightarrow N$) of $q$ eigenvectors $U_k$
 ($k=1 \rightarrow q$) can be defined as 
 
\begin{eqnarray}
 P_{Nq}(Z_1,Z_2,..Z_q, e_1,..,e_q; Y)= \int\; \prod_{k=1}^q { \delta_k}\;  P_{ef,ev}(\lambda_1,..\lambda_N, U_1,..,U_N) \; \prod_{j=1}^N  {\rm D}\tau_j, 
\end{eqnarray}
with symbol $\delta_k$  defined in eq.(\ref{delk})  and ${\rm D}\tau_j \equiv  {\rm d}\lambda_j \;  {\rm D}^{\beta} U_j$. 
 Proceeding exactly as above for $P_{N1}$, the $Y$-governed diffusion of $P_{Nq}$ 
can be shown to be described as   
    
 \begin{eqnarray}
 {\partial P_{Nq}\over \partial Y} &=&  \sum_{k=1}^q 
\left[ {\tilde F}_{k} + {\tilde F}_{k}^* +{L}_{e_k}P_{Nq}  \right] 
\label{dpnq}
\end{eqnarray}
with
\begin{eqnarray}
{\tilde F}_k &=& {\mathcal F}_{k} - {\beta^2 \over 2^{\nu+2} } \sum_{l=1;\not= k}^q \; 
{(e_k+e_l)^{\nu} \over (e_k-e_l)^2} \; \sum_{m=1}^N  {\partial^2  ( z_{nk} z_{ml} P_{Nq}) \over \partial z_{mk} \partial z_{nl}} 
\label{tfk}
\end{eqnarray}
Here  ${\mathcal F}_{k}$ is same as  $F_k$ in eq.(\ref{fk}) but with 
following replacements: $P_{N1} \rightarrow P_{Nq}$, $Q_{mn;k}^{rs} \rightarrow {\mathcal Q}_{mn;k}^{rs}$. For clarity purposes, we rewrite it here:
\begin{eqnarray}
{\mathcal F}_k &=& {\beta^2\over 4} \; \left[\sum_{m=1}^N{\partial \over \partial z_{mk}} \left[  z_{mk} {\mathcal Q}_{mm;k}^{02} \right]
+ \sum_{m,n=1}^N{\partial^2 \over \partial z_{mk} \partial z^*_{nk}} {\mathcal Q}_{mn;k}^{12} \right], 
\label{fkm}  
\end{eqnarray}
 Here
\begin{eqnarray}
{\mathcal Q}_{mn;k}^{rs} 
&=& 2^{-\nu} \; \sum_{j \not=k} \int \left( \prod_{n=1}^q { \delta_n} \right) \; {(e_k+e_j)^{\nu}\over (e_k-e_j)^s} \;(z_{nj} z^*_{mj})^r 
\; P_{ef,ev}(\lambda_1,..\lambda_N, U_1,..,U_N) \; \prod_{l=1}^N  {\rm D}\tau_l 
\label{qqr1}  \nonumber \\
&=& {1\over 2^{\nu}} \sum_{l=1; l\not=k}^q  {(e_k+e_l)^{\nu} \over (e_k-e_l)^s} \; (z_{nl} z^*_{ml})^r \; P_{Nq}  + {1\over 2^{\nu}}  \; \sum_{j; j >q} \int {(e_k+e_j)^{\nu} \over (e_k-e_j)^s} \;(z_{nj} z^*_{mj})^r \;P_{N(q+1)} \;  {\rm D}\tau_j. \nonumber \\
\label{q2}
\end{eqnarray}
with $P_{N(q+1)} \equiv P_{N(q+1)}(Z_1,Z_2,..,Z_q, Z_j)$ is the joint probability density of $q+1$ 
eigenfunctions, namely,  $Z_1,Z_2,..,Z_q$ along with $Z_j$ (with $j>q$). 
 Considering a similar approximation as in the case of $Q_{mn;k}^{rs}$, one can write
\begin{eqnarray}
{\mathcal Q}_{mn;k}^{r2} \approx  {1\over 2^{\nu}} \; \sum_{l=1; l\not=k}^q  {(e_k+e_l)^{\nu}\over (e_k-e_l)^2} \; (z_{nl} z^*_{ml})^r  \; P_{Nq} +  {\mathcal K}_2 \; \left(\overline{\langle z_{nk} z_{mk}^*\rangle} \right)^r \; P_{Nq}.
\label{qnq}
\end{eqnarray} 

Similarly $L_{e_k}$ is again given  by eq.(\ref{lek})  but with $P_{N1}$ replacing $P_{Nq}$ and  $T_q$ replacing ${T}_1$ where 
\begin{eqnarray}
{T}_q(e_k)  &=&  e_k^{\nu} \; \sum_{j; \not=k} \int \left( \prod_{n=1}^q { \delta_n} \right) \; {P_{ef,ev} \over (e_k-e_j)}
 \; \prod_{l=1}^N  {\rm D}\tau_l    \nonumber \\
& \approx & {\mathcal K}_1 \; P_{Nq} + \sum_{n=1; n \not=k}^q {e_k^{\nu} \; P_{Nq} \over e_n -e_k}.
\label{tq}
\end{eqnarray} 
The above along with eqs.(\ref{tfk},\ref{fkm},\ref{qnq}) reduces eq.(\ref{dpnq}) in a closed form differential equation for $P_{Nq}$. 
Note for the Gaussian case ($\nu=0$), eq.(\ref{dpnq}) is same as eq.(25) of \cite{pswf} (with misprints in eq.(26) of \cite{pswf} corrected here).


%
 
 
 \section{Fluctuation Measures of Eigenfunctions}

Eqs.(\ref{pp1}, \ref{dpn1}, \ref{dpnq}) describe the evolutions of the joint probability densities of various combinations of the eigenfunction components and corresponding eigenvalues. 
%
Following similar steps as used in \cite{pswf}, one can again derive the diffusion equations for various fluctuation measures but presence of new  terms of type $(e_k+e_l)^{\nu}$ for Wishart case is expected 
to increase the technical complexity of  the partial integrations applied at various stages. Here we consider some relevant examples.


\subsubsection{Local eigenfunction intensity}
      
The local eigenfunction intensity $u$ is an important measure of the influence of a specific basis state on the wavefunction dynamics \cite{mir}. Its distribution can be defined as 
\begin{eqnarray}
P_u(u; e) = \langle {1\over N} \sum_{k=1}^N \delta(u-N |z_{nk}|^2) \delta(e-e_k) \rangle. 
\label{pu0}
\end{eqnarray}
To derive the $Y$-dependence of $P_u(u)$, we first 
 consider the distribution 
$P_{11}(x,e)$  of an eigenfunction component $x=N^{1/2} z_{nk} =(u^{1/2})$ at an 
energy $e$, defined as  
 \begin{eqnarray}   
 P_{11} (x,x^*,e; Y)= {1\over N} \sum_{k=1}^N  \langle  \;  \delta^{\beta}_x \; \delta_e \rangle 
= \int \delta^{\beta}_x \; \delta_e \;  P_{1N}(Z,E,Y) \; {\rm D}\tau 
\end{eqnarray}
 where $\delta^{\beta}_x= \delta(x- \sqrt{N} z_{nk}) 
\delta^{\beta-1}(x^*- \sqrt{N} z^*_{nk})$ and $\delta_e= \delta(e-e_k)$ 
and ${\rm D}\tau \equiv \prod_{j=1}^N \; {\rm d}e_j \; {\rm d}^{\beta}Z_j$. Using eq.(\ref{pp1}) and following 
the same steps as used in section IV. A of \cite{pswf}), 
the diffusion equation for $P_{11}(x,e)$ can be expressed in the same form  as in the  Gaussian case (see eq.(29) of \cite{pswf}) 
\begin{eqnarray}
 {\partial P_{11} \over\partial Y}=
{\beta^2 \over 4 }\left[ 2{\partial^2 G_{1} \over\partial x \partial x*}
 +  {\partial (x G_{0}) \over\partial x} + 
    {\partial (x^* G_{0}) \over\partial x^*}\right]
 + L_e P_{11}
\label{p11}
 \end{eqnarray}
but now 
\begin{eqnarray}
 G_{r}(x,e) \equiv   {N^r \over 2^{\nu} } \; \sum_{j;j\not=k}\int \delta^{\beta}_x \; 
 \delta_e \; { (e_k+e_j)^{\nu} \over (e_k-e_j)^{2} } \; |z_{nj}|^{2r}
 P_{1N} \; {\rm D}\tau, 
 \end{eqnarray}
with $r=0,1$ and 
\begin{eqnarray}
 L_e P_{11} &=& \int \delta^{\beta}_x\; \delta_e\; [L_E P_{1N}]\; {\rm D}\tau , \label{le0} \\
&=&  {\partial \over \partial e}\left[
 \beta \; a(e) \; P_{11} + {\partial  \over \partial e} (e^{\nu} \; P_{11}) + \beta \;   T_0(e) \right]
\label{lee}
\end{eqnarray}
where $a(e)$ is given by eq.(\ref{ae}) and 
\begin{eqnarray}
T_0(e) = e^{\nu} \; \sum_{j \not=k} \int \delta^{\beta}_x\; \delta_e\; { P_{1N} \over (e_j-e_k)}  \; {\rm D}\tau  \quad \approx  \quad  {\mathcal K}_1 \; \; P_{11};
\end{eqnarray}
with ${\mathcal K}_1$ given by eq.(\ref{kk}). 
Here the 2nd equality follows under similar approximation as in eq.(\ref{tk1}).
 
Eq.(\ref{p11}) describes the sensitivity of the local intensity distribution to the energy 
scale  $e$ as well as perturbation  parameter $Y$. 
Using the same approximation as in the case of $Q_{mn;k}^{rs}$ (see appendix C),  $G_0$ can be approximated as
\begin{eqnarray}
G_{0} &\approx &  {\mathcal K}_2 \; \int \;  \delta^{\beta}_x \; \delta_e \; P_{1N} \; {\rm d}\tau  = {\mathcal K}_2   \; P_{11}(x,e),
\label{g1}
\end{eqnarray} 
with ${\mathcal K}_2$ given by eq.(\ref{kk}).
Similarly $G_1$ can be reduced as 
%
%
\begin{eqnarray}
G_1 
&\approx &  {\mathcal K}_2    \; \overline {\langle |x|^2 \rangle} \;  P_{11}(x; e).  
\label{g2} 
\end{eqnarray} 
with $\overline{\langle |x|^2 \rangle}$ implying a local spectral as well as ensemble average defined as $\overline{\langle |x|^2 \rangle} = {1\over E_c} \; \int_{E_c}  \langle |x|^2 \rangle \; {\rm d}e$. 

%
%
With help of  eq.(\ref{p11}), eq.(\ref{g1}) and eq.(\ref{g2}), the evolution equation for $P_u(u; e)$ can now be derived as follows. Using the definition  
$P_u(u; e) =  \int \delta(u -|x|^2) \; P_{11}(x,x^*,e) \; {\rm d} x \; {\rm d}^{\beta-1}x^*$, $ {\partial P_u \over\partial Y}$ can 
be expressed in terms of  an integral over ${\partial P_{11} \over\partial Y}$. Subsequent substitution of eq.(\ref{p11}) and partial integrations  
then lead to   $Y$-governed  evolution equation for $P_u(u,e; Y)$ 
\begin{eqnarray}
 {\partial P_u \over\partial Y}= 2 \; {\mathcal K}_2 \;  \left[ \overline{\langle u \rangle} \; \; 
   {\partial^2 ( u P_u) \over\partial  u^2}  
 + {\beta \over 2} \; {\partial \over\partial u} \left(u -\overline{\langle u \rangle}\right)P_u \right] +  L_e P_u 
\label{pu}
 \end{eqnarray}
with $\overline{\langle x^2 \rangle}$ now written as $\overline{\langle u \rangle}$ and
\begin{eqnarray}
L_e = {\partial \over \partial e}\left[
 \beta \; a(e)  + \beta \; {\mathcal K}_1 + {\partial  \over \partial e} e^{\nu}  \right]
\label{le}
\end{eqnarray}

The energy-dependence of eq.(\ref{pu}) indicates a non-stationary behavior of the local intensity distribution. Again in the stationary limit ${\partial P_u \over\partial Y} \to 0$ or $Y \to \infty$, it is easy to check that $\int P_u \; {\rm d}e$ satisfies the Porter-Thomas distribution 
for the local intensity $P_u(u) \propto u^{\beta/2-1} \; {\rm e}^{-\beta u/2}$
which  describes the stationary Wishart cases in infinite size limit \cite{me}; 
(This is as expected because  the ensemble density $\rho(L)$ approaches WOE or WUE  in $Y \to \infty$ limit).

Eq.(\ref{pu}) can further be used to calculate ensemble averaged local intensity at an energy $e$, defined as $\langle u(e) \rangle ={1\over R_1(e)} \int  u \; P_u(u; e) \; {\rm d} u$ with $R_1(e)$ as the ensemble averaged level density (see Sec.IV). Multiplying eq.(\ref{pu}) by $u$ and integrating gives 
\begin{eqnarray}
 {\partial  ( R_1 \langle u \rangle) \over\partial Y}=- \; \beta \; R_1 \; {\mathcal K}_2  \;  \left(\langle u \rangle - \overline{\langle u \rangle}  \right) +   \; L_e (R_1 \; \langle u \rangle). 
\label{pu1}
 \end{eqnarray}
where $R_1(e,Y)$ is given by the equation ${\partial  R_1 \over\partial Y} = L_e R_1$. 
(It is easy to check that eq.(\ref{pu1}) gives correct result in stationary limit ${\partial (R_1 \overline{\langle u \rangle}) \over\partial Y} \to 0$ or $Y \to \infty$: as eigenvalues and eigenfunctions are independent in this limit,  we have $L_e (R_1 \langle u \rangle) = \langle u \rangle \; L_e R_1(e) =0$. Further the stationarity implies  $\langle u \rangle =\overline{\langle u \rangle}$.  This gives right side of eq.(\ref{pu1})  zero as expected).





As clear from the above, $\langle u(e) \rangle$ undergoes a $Y$-governed  dynamics in the $e$-space. 
%
 For regions with a slow variation of $R_1$ with respect to $e$,  
eq.(\ref{pu1}) can further be simplified by  considering a local spectral averaging  of $\langle u \rangle$ defined as 

\begin{eqnarray}
\overline{\langle u \rangle} ={\int_{e-D_e}^{e+D_e} \; {\rm d}e \int \; {\rm d}u \; u \; P_u(u; e) \over \int_{e-D_e}^{e+D_e} \; {\rm d}e \int \; {\rm d}u \; P_u(u; e) }
={R_1(e) \over N_e} \; \int_{e-D_e/2}^{e+D_e/2} \; {\rm d}e  \; \langle u(e) \rangle 
\label{uae}
\end{eqnarray}
with $N_e$ as the number of levels used for local averaging: $N_e= \int_{e-D_e}^{e+D_e} \; R_1(e) {\rm d}e$. 
%
This permit one to replace  $\langle u \rangle$ by $\overline{\langle u \rangle}$ and  leads to 
\begin{eqnarray}
 {\partial \overline{\langle u \rangle} \over\partial Y}=
\beta   \;  {\partial \over \partial e} \left( a(e) + {\mathcal K}_1 \right) \; \overline{\langle u \rangle}  +    {\partial^2  \over \partial e^2} \; e^{\nu} \; \overline{\langle u \rangle}, 
\label{pu1p}
 \end{eqnarray}
where $a(e)$ and ${\mathcal K}_1$ are given by eq.(\ref{ae}) and eq.(\ref{kk}).
A solution of the above equation for Wishart BE ($\nu=1)$ has an exponential decay with $e$:
\begin{eqnarray}
\langle u(Y) \rangle = u_0(Y) \; {\rm exp}\left[-{\beta_2  \; e\over (1-q)} \right]
\label{pus}
 \end{eqnarray}
where $\beta_2=\beta \left( \left({2\over \beta} \right)^{\nu} +{2 \nu {\mathcal N} \over E_c}\right)$ and 
$u_0(Y)= {1\over q} \; \left( {q\over 1-q}\right)^{(\beta a_0+2)}$ with $q= v_0 \; {\rm e}^{- \beta_2 (Y-Y_0)}$ with $v_0$ given by initial conditions. 
Note the solution of eq.(\ref{pu1p}) for Gaussian case ($\nu=0$) has a Gaussian decay with $e$: 
\begin{eqnarray}
\langle u \rangle = {\langle u_0 \rangle \over \sqrt{1-q^2}} \; {\rm exp}\left[-{(\tilde{e}- q \; \tilde {e_0})^{2} \over (1-q^2)} \right]
\label{pusg}
 \end{eqnarray}
with  $\tilde{e}=e+{2 N\over E_c}$, $\tilde{e}_0=e_0+{2 N\over E_c}$ and $\langle u_0 \rangle$ as the initial intensity at $e=e_0$ and $Y=Y_0$.

\subsubsection{Inverse participation ratio}

The moments of the eigenfunction intensity, also known as inverse participation ratios  are standard tools to measure the spread of an eigenfunction in the basis-space.  For an eigenfunction $Z_k$ in a discrete basis, it can be defined as
$I_q(k)=\sum_{j=1}^N |z_{jk}|^{2q}$   and its ensemble average can be written as 
\begin{eqnarray} 
\langle  I_q(e) \rangle ={N^{1-q}} \; \int_0^{\infty} u^{q} \; P_u (u, e) \; {\rm d} u 
\label{iqa}
\end{eqnarray}.

The $Y$-governed evolution equation for $\langle I_q \rangle$ for an arbitrary eigenfunction $Z_k$ can be derived from eq.(\ref{pu}) as follows.  
By differentiating  eq.(\ref{iqa}) with respect to $Y$ and using eq.(\ref{pu}), we obtain 
 
 \begin{eqnarray}
{\partial \langle  I_q \rangle  \over\partial Y} 
\approx \beta  \; {\mathcal K}_2 \; q \; \left(t_1 \; \langle  I_{q-1}\rangle -  \langle I_q \rangle \right) +   \; L_e \;  \langle  I_q \rangle,  
\label{iqa1}
 \end{eqnarray}
where $L_e $  given by eq.(\ref{le}) and
$t_1(q) =\left(1 + {2(q-1) \over \beta  }\right) \; {\overline{\langle u\rangle}  \over N}$.
%
%


Eq.(\ref{iqa1}) describes the variation of $\langle I_q \rangle $ with respect to energy $e$ and parameter $Y$. As in the case of $\langle u \rangle$, it can further be simplified by a local  spectral averaging.  For $e$-regions with almost constant level density $R_1(e)$, the evolution of $\overline{\langle  I_q \rangle} =  {1 \over N_e} \int_{e-D_e}^{e+D_e}  \; \langle  I_q \rangle \; R_1(e) \; {\rm d}e \approx {R_1(e) \over N_e} \int_{e-D_e}^{e+D_e}  \; \langle  I_q \rangle \; {\rm d}e $  can  be described as 

\begin{eqnarray}
{\partial \overline{\langle I_q \rangle} \over\partial \Lambda_I}= \left(t_1 \; \overline{\langle  I_{q-1}\rangle} - t_2 \overline{\langle I_q \rangle}  \right) + {1 \over q \beta {\mathcal K}_2} \;  \Gamma_e \; \overline{\langle I_q \rangle} 
\label{iu1}
 \end{eqnarray}

%
%
where $\Gamma_e =  \left(\left({2\over \beta} \right)^{\nu} \; e + {2 N\over E_c} e^{\nu} + b_0 \right)  {\partial  \over \partial e} +e^{\nu} \; {\partial^2 \over \partial e^2} $ with $b_0= {\nu \over 2} (N-N_a+3)$ and ${\mathcal K}_2$ given by eq.(\ref{kk}), $\Lambda_I = {q \; \beta \;  {\mathcal K}_2 \; (Y-Y_0)}$, 
$t_2(q) =
1+{ 1\over q  \; {\mathcal K_2}} \; \left[\left({2\over \beta} \right)^{\nu} + {2 \nu  N\over E_c} \right]$.



The average local intensity $\overline{\langle u \rangle}$ and therefore $t_1(q)$ is a function of  both $e$ as well as $Y-Y_0$. Although the last term in eq.(\ref{iu1}) can be neglected due to ${\mathcal K}_2 \sim O(N^s)$ with $s >1$ , it still retains an $e$-dependence  through  $\Lambda_I$ as well as $t_1$. For cases in which  variation of $t_1$ with respect to $Y$ is negligible,  a solution of eq.(\ref{iu1}) can be written in form of a recurrence relation 

\begin{eqnarray}
\overline{\langle  I_q(\Lambda_I)\rangle} \approx  {\rm e}^{-t_2 \; \Lambda_I}\left[\overline{\langle I_q(0)\rangle} + t_1 \int_0^{\Lambda_I} \overline{\langle  I_{q-1}(r) \rangle} \; 
{\rm e}^{t_2 \; r} \; {\rm d}r \right]. 
\label{iqao}
\end{eqnarray}


For a finite but large $\Lambda_I$, $I_q$ for first few $q$ values can be given as 
\begin{eqnarray}
\overline{\langle I_0(\Lambda_I) \rangle} &=& {t_2(1) \over t_1(1)},  \qquad \quad
\overline{\langle I_1(\Lambda_I) \rangle} = 1, \\
\overline{\langle  I_2(\Lambda_I)\rangle} &=& {t_1(2) \over  t_2(2)} +\left(I_2^{(0)} - {t_1(2) \over t_2(2)}   \right) {\rm e}^{- t_2 \Lambda_I}  \\
\overline{\langle  I_3(\Lambda_I)\rangle} &=& {t_1(3) \over  t_2(3)}   {t_1(2) \over  t_2(2)}   + \left(I_2^{(0)}  - {t_1(2) \over  t_2(2)}  \right) {t_1(3) {\rm e}^{-t_2(2)\Lambda_I}  \over t_2(3)-t_2(2)} + \nonumber \\
& & \left( I_3^{(0)} - {t_1(3) \over  t_2(3)}   {t_1(2) \over  t_2(2)}   
-   \left(I_2^{(0)} -{t_1(2)\over t_2(2)} \right) {t_1(3) \over t_2(3)-t_2(2)}  \right)  \; {\rm e}^{- t_2(3) \Lambda_I} 
\label{iqap}
\end{eqnarray}
where  $I_q^{(0)} \equiv \overline{\langle  I_q(0) \rangle}  $. As $q$ increases, the number of terms in the  expression for $I_q$  
 increase rapidly. For large $q$ or large $\Lambda_I$, it can however be approximated as 

 \begin{eqnarray}
\overline{\langle  I_q(\Lambda_I)\rangle} \approx    \prod_{k=2}^{q} \; {t_1(k)\over t_2(k)} +O({\rm e}^{-t_2 \Lambda_I} )
\label{iqaq}
\end{eqnarray}

In the limit $\Lambda_I \to \infty$ (which corresponds to a stationary Gaussian or Wishart ensemble e.g GOE, GUE, WOE, WUE),  ${\overline\langle u \rangle}\sim 1$ which implies $t_1 \to {2(q-1)+\beta \over N}$ and $t_2 \to 1$ (latter following as ${\mathcal K}_2 = {2 N \over E_c^2} \; {\rm e}^{\nu}$ with $E_c \sim \Delta_e$. The large $\Lambda_I$-limit of   $ \langle  I_q(\Lambda_I)\rangle$ is then in agreement with expected stationary limit:  $ \langle  I_q(\Lambda_I)\rangle \approx {(2q-1)!! \over N^{q-1}}$ for $\beta=1$ and $ \langle I_q(\Lambda_I)\rangle \approx {q! \over N^{q-1}}$ for $\beta=2$ \cite{mir}.

Many localization to delocalization studies indicate the probability density $P_I(I_q)$ of $I_q$ as an important measure for the wavefunction fluctuations. The diffusion equation for $P_I(I_q)$ for the Gaussian case was derived in \cite{pswf}. Following similar steps, It can be derived 
for the Wishart case too. 




\subsubsection{Correlation between two wavefunctions at different energies}

An important measure to describe  the eigenfunction localization,  the two-point correlation $C(e',e")$ between two eigenstates, say $Z_a$ and $Z_b$ with eigenvalues $e, e'$ respectively, can be defined as    
\begin{eqnarray}
C(e', e'') =\sum_{a, b \atop a\not=b} \sum_{m=1}^N |z_{ma} |^2 \; |z_{mb}|^2 \; \delta(e'-e_a) \delta(e''-e_b)
\label{corr}
\end{eqnarray}
 (with $z_{ma}$ as the $m^{th}$ component of the eigenfunction $Z_a$).  Its ensemble average can be expressed in terms of the joint probability density $P_{N2}(Z_a, Z_b, e_a, e_b)$ of $Z_a$ and $Z_b$ and corresponding eigenvalues: 
\begin{eqnarray}
\langle C(e',e'') \rangle = \sum_{a,b; a\not=b} \; \langle \; C_{ab} (Z_a, Z_b, e', e'') \; \rangle
\label{corr1}
\end{eqnarray} 
where $C_{ab}=\sum_{k=1}^N  |z_{ka} |^2 \; |z_{kb}|^2 \; \delta(e'-e_a) \; \delta(e''-e_b)$   and   $ \langle C_{ab} \rangle = \int  \; C_{ab} \; P_{N2} \; {\rm D}Z_a \; {\rm D}Z_b \; {\rm d}e_a \; {\rm d}e_b$.

As intuitively expected, an ensemble averaged $C(e',e'')$ is related to the 2-point spectral correlation $R_2(e',e")$; this in turn connects the criticality criteria in the eigenfunction statistics to that of eigenvalues \cite{pswf, psand}.  For example, $\langle C \rangle = C_0 \; R_2(e',e'')$ for completely delocalized eigenfunctions with $|z_{ka} |^2, |z_{kb} |^2 = 1/N$ and $C_0=1$; the fluctuations in the eigenfunction components however  result in a change of $C_0$. 

In general, a parametric diffusion of eigenfunctions leads to diffusion of $\langle C_{ab}(e',e'') \rangle$

\begin{eqnarray}
{\partial \langle C_{ab} \rangle \over \partial Y}= \int C_{ab} \; {\partial P_{N2}\over \partial Y} \; {\rm D}\tau_a \; {\rm D}\tau_b
\label{cee}
\end{eqnarray}
where ${\rm D}\tau_a \; {\rm D}\tau_b \equiv DZ_{a} \; DZ_{b} \; {\rm d}e_a \; {\rm d}e_b$. 
Substitution of eq.(\ref{dpnq}), with $q=2$ while using $Z_a, Z_b, e_a, e_b$ instead of $Z_1, Z_2, e_1, e_2$, in the above equation leads to 
\begin{eqnarray}
{\partial \langle C_{ab} \rangle \over \partial Y}=  \left(J_1 +J_1^* + J_2 +J_2^*+ J_3\right)
\label{cee1}
\end{eqnarray}
where
\begin{eqnarray}
J_1 &=&   \;\int \; C_{ab} \; 
\left[ {\mathcal F}_a + {\mathcal F}_b \right] \; {\rm D}\tau_a \; {\rm D}\tau_b  \label{j1}\\
J_2 &=&  -{\beta^2 \over 2} \int \; C_{ab} \;{(e_a+e_b)^{\nu}\over (e_a-e_b)^2} \; \sum_{m,n=1}^N  {\partial^2  ( z_{ma} z_{nb} P_{N2}) \over \partial z_{na} \partial z_{mb}}  \; {\rm D}\tau_a \; {\rm D}\tau_b \label{j2}\\
J_3 &=& \int \; C_{ab} \; \sum_{k=a,b}  {\partial \over \partial e_k} \left[
  \beta \; a(e_k)  + \beta \; e_k^{\nu}  \; T_2(e_k) +  {\partial \over \partial e_k} \; e_k^{\nu} 
\right] \; P_{N2} \; {\rm D}\tau_a \; {\rm D}\tau_b
\label{j3}
\end{eqnarray}
where ${\mathcal F}_a$ is same as ${\mathcal F}_1$ given by eq.(\ref{fkm})  with $q=2$ and  $Z_a, Z_b, e_a, e_b$ replacing $Z_1, Z_2, e_1, e_2$: 
\begin{eqnarray}
{\mathcal F}_a &=& {\beta^2 \over 4 } \sum_{m, n=1}^N   {\partial^2 \over \partial z_{na} \partial z^*_{ma}} \; {\mathcal Q}_{mn;a}^{12}
+  {\beta^2 \over 4 }  \sum_{n=1}^N {\partial \over \partial z_{na}} \; \left( z_{na}{\mathcal Q}_{nn;a}^{02}  \right)
\label{fka}
\end{eqnarray}
and
\begin{eqnarray}
{\mathcal Q}_{mn;a}^{r2} 
&\approx &  {1\over 2^{\nu}} \; {(e_a+e_b)^{\nu} \over (e_a-e_b)^2} \; (z_{nb} z^*_{mb})^r \; P_{N2}  + 
{\mathcal K}_2 \; 
\left( \overline{\langle z_{na} \; z_{ma}^* \rangle} \right)^r P_{N2}.
\label{q5}
\end{eqnarray}
%
Similarly ${\mathcal F}_b$ and  ${\mathcal Q}_{mn;b}^{rs} $ can also be given by the above equations by replacing $a \to b$ everywhere.


Applying partial integration repeatedly, $J_1$ can be rewritten as 
\begin{eqnarray}
J_1 &=& - {\beta\over 2} \; \sum_{k=1}^N \int |z_{ka}|^2 \; |z_{kb}|^2  \; \left( {\mathcal Q}_{kk;a}^{02} + {\mathcal Q}_{kk;b}^{02} \right) \; \delta_a \; \delta_b \; {\rm D}\tau_a \; {\rm D}\tau_b  \nonumber \\
&+& {\beta \over 2} \; \sum_{k=1}^N  \int \left( |z_{kb}|^2  {\mathcal Q}_{kk; a}^{12} + |z_{ka}|^2  {\mathcal Q}_{kk;b}^{12} \right) \; \delta_a \; \delta_b  \; {\rm D}\tau_a \; {\rm D}\tau_b
\label{jj1}
\end{eqnarray}
Using the approximations (\ref{q5}), eq.(\ref{jj1}) can further be reduced as
\begin{eqnarray}
J_1 = - {\beta\over 2^{\nu+1}} \;  {(e'+e'')^{\nu}\over (e'-e'')^2} \; \left[ 2 \; \langle C_{ab} \rangle   - B_{e' }- B_{e''} \right]
\label{jk1}
\end{eqnarray}
where 
\begin{eqnarray}
B_{e'}  &=&   \int I_{2a}  \; \delta(e'-e_a) \; \delta(e''-e_b) \; P_{N2} \; {\rm D}\tau_a \; {\rm D}\tau_b. 
\label{be}
\end{eqnarray} 
and $I_{2a}=\sum_{k=1}^N \;  |z_{kb}|^4$ is the inverse participation ratio for the eigenfunction $Z_a$ at energy $e_a$.
Similarly, by replacing $I_{2a} \to I_{2b}$, the above equation gives  $B_{e''}$.  
Note, $B_{x} \approx \langle I_{2x} \rangle \; R_2(e', e'')$ with $x=e', e''$  which follows by replacing $I_{2a}, I_{2b}$ by their ensemble average values $\langle I_{2e'} \rangle$ and $\langle I_{2e''} \rangle$ at the energies $e', e''$, respectively.

To calculate $J_2, J_3$, we again partial integrate eq.(\ref{j2}) and eq.(\ref{j3})  which gives  
\begin{eqnarray}
J_2 =  - 2^{1-\nu}  \; {(e'+e'')^{\nu} \over (e'-e'')^2} \;  \langle C_{ab} \rangle 
\label{jj2}
\end{eqnarray}
and 
%
\begin{eqnarray}
J_3 &=& \sum_{k=a,b}  \int \left[ {\partial^2 C_{ab} \over \partial e_k^2} \; e_k^{\nu} \; P_{N2} - {\partial C_{ab} \over \partial e_k} \; \left(\beta \; a( e_k) \; P_{N2} + \beta  \; T_2(e_k) \right)\right] \; {\rm D}\tau_a \; {\rm D}\tau_b  \\
&=& \sum_ {x=e', e''} \left[{\partial^2    \over \partial x^2} \; x^{\nu}   +\beta \; {\partial  \over \partial x} \;\left( a(x) + 
{N\over E_c} \; x^{\nu}  \right) \right] \langle C_{ab}\rangle 
+\beta \left({\partial \over \partial e'} e'-{\partial \over \partial e''} e'' \right) {\langle C_{ab} \rangle \over (e''-e')} \nonumber \\
 \label{jj3}
\end{eqnarray}
%
Substitution of eqs.(\ref{jk1}, \ref{jj2}, \ref{jj3}) in eq.(\ref{cee1}) now leads to the $Y$-governed evolution of $\langle C_{ab}(e,\omega) \rangle$, with $e'=e+\omega, e''=e-\omega$,  which on summing over $a,b$ leads to

\begin{eqnarray}
{\partial  \langle C \rangle \over\partial Y} &\approx & \left[{1\over 2} \left( {\partial^2    \over \partial e^2} +  {\partial^2    \over \partial \omega^2} \right) e^{\nu}  +  \nu \; {\partial^2   \; \omega  \over \partial e \; \partial \omega} 
+ \beta \; {\partial  \over \partial e} \; \left(\left({2\over \beta}\right)^{\nu}  e + {N\over E_c}  \; e^{\nu} +a_0 -{\nu \over 2} \right) \right]\langle C \rangle  + 
\nonumber \\
& + &  \beta {\partial  \over \partial \omega} \; \left(  \left({2\over \beta}\right)^{\nu} \omega +  {\nu \; N\over E_c} \; \omega -{e^{\nu} \over 2 \omega} \right) \langle C \rangle  - (\beta+2) {e^{\nu} \over 4 \omega^2}  \langle C \rangle 
+ \nonumber \\
&+&  { \beta \; e^{\nu} \over 8 \omega^2} (\langle I_{2(e+\omega)} \rangle + \langle I_{2(e-\omega)}\rangle) \; R_2(e,\omega), 
\label{cee2}
\end{eqnarray}
with $a_0 = \nu (N-1-N_a)/2$.
%
As clear from the above, the correlation between two eigenfunctions at different energies varies along the energy axis.  
Furthermore the energy-dependence of the correlation is different for Gaussian ($\nu=0$) and Laguerre Brownian ensemble ($\nu=1$). 
For locally stationary correlation in energy  i.e those for which a variation with respect to $e$ can be ignored, a substitution of ${\partial \langle C \rangle \over \partial e} \approx 0$ reduces eq.(\ref{cee2}) to

                                                                                                                                                                                                                                                                            \begin{eqnarray}
2 \; {\partial  \langle {C} \rangle \over\partial \Lambda_e} &\approx&\left[  {\partial^2     \over \partial r^2}    + \beta  {\partial \over \partial r} \left(2\eta r +{1\over r} \right) -  {(\beta+2)\over 2 \; r^2} + 2 \beta \eta \right] \; \langle {C} \rangle + { \beta \over  4 \; r^2}  \; \langle I_{2(r_0+r)} + I_{2(r_0-r)}\rangle  \; { R}_2(r_0, r)  \nonumber \\
\label{cee3}
\end{eqnarray}
where $r_0, r$ are the rescaled energies $e = r_0  \; \Delta_e, \omega = r \; \Delta_e$ with $\Delta_e$ as the local mean level spacing, $\Lambda_e$ is defined in eq.(\ref{alm1}) and $\langle {C} \rangle$ and $R_2$ are redefined as $\frac{\langle {C} \rangle}{ \Delta_e^2} \to \langle {C} \rangle$, $\frac{R_2(e, \omega)}{ \Delta^2(e)} \to R_2(r_0, r)$ and $\eta= {\rm e}^{-\nu} \; \Delta_e^2 \; \beta_2$ with $\beta_2= \left(\left({2\over \beta}\right)^{\nu} + {\nu  N \over E_c} \right)$.  (Here the terms containing $\Delta_e$ are neglected due to being $o(1/N)$ smaller as compared to other terms). 
%

As a check, let us first consider the stationarity limit ${\partial  \langle  C \rangle \over\partial \Lambda_e}=0$ or, alternatively, the limit $\Lambda_e \to \infty$ which corresponds to stationary ensembles  with delocalized eigenfunctions: using $\langle C \rangle = C_0 \; R_2(r_0, r)$ and eq.(\ref{rn}) 
for $R_2$ (with ${\partial  R_2 \over\partial \Lambda_e}=0$ and neglecting the integral term for small $r$),  it is easy to check that $C_0 \approx  { \beta \over  (\beta+2)} \; \langle I_{2 r_0} \rangle $; (here the eigenfunctions statistics being energy independent, $\langle I_{2(r_0+r)} \rangle = \langle I_{2(r_0-r)} \rangle=\langle I_{2 r_0} \rangle$).

The next desirable step would be to solve eq.(\ref{cee3}). Noting its singularity at $r=0$, a solution for small-$r$ can be obtained by a   Taylor's series expansion of 
$ \langle I_{2(r_0 \pm r)} \rangle$ and  $\langle {C} \rangle$ in the neighbourhood of $r=0$: 
$\langle {C} \rangle = r^s \; \sum_{n=0}^{\infty} \; d_n(\Lambda_e) \; r^n $. Clearly $s, d_0, d_1$ depend on the small-$r$ behavior of $R_2(r; \Lambda_e)$. Expanding $R_2(r, \Lambda_e)$ in Taylor's series around $r=0$  as $R_2 =  r^{p} \sum_{m=0}^{\infty} v_m (\Lambda_e) \; r^m$ and approximating $\langle I_{2, r_0+r} \rangle + \langle I_{2, r_0-r}\rangle) \approx 2\langle I_{2, r_0} \rangle$, one has two possible solution for $s$: (i) $s=p$, $ d_0={-4 \beta \langle I_{2 r_0} \rangle \over (p-1)(p-\beta) -(\beta+2)}.
, d_1=0$ and (ii) $s=p-1,  d_0=0, d_1={-4 \beta \langle I_{2 r_0} \rangle \over (p-1)(p-\beta) -(\beta+2)}$. This reveals the connection between $2$-point wavefunction correlation with $2$-point spectral-density correlation: $\langle {C} \rangle \approx R_2(r)$ for small-$r$.  
Higher $d_n$ ($n >1$) are given by the hierarchic relation: $2 \; {{\rm d} d_{n-2} \over {\rm d}\Lambda_e} = \left[ (n+s - \beta) (n +s -1)  - 2^{\nu-1} \; (\beta+2) \right] \; d_n$.


\section{Conclusion}

We conclude with a summary of our main results: based on a  combination of 2nd order perturbation theory  for Hermitian matrices and a Markovian dynamics of matrix elements, we have  analytically derived the moments for the eigenfunction components for a Wishart Brownian ensemble.  This in turn leads to a diffusion equation for the probability densities of various eigenfunction fluctuation measures; here we have explicitly derived  the equations for the distribution of the components of a single eigenfunction, of a given component of all eigenfunctions and that of all components of many eigenfunctions along with their eigenvalues. The equations are later on applied to derive the parametric dependence of the standard fluctuation measures i.e local eigenfunction intensity, inverse participation ratio as well as eigenfunction correlations at two different energies. The well-known applications of these measures to the studies based on eigenfunctions dynamics \cite{mir} and their accessibility  for the  experimental/numerical analysis makes our results relevant for a wide-range of complex systems which can be modeled by the non-equilibrium Wishart ensembles.  

In this paper, our main focus has been on the derivation of the evolution equations for various eigenfunction measures begining from arbitrary initial conditions. The complexity of the equations makes their solution  technically complicated.  But, based on the initial conditions,  approximate solutions can be obtained  which provide relevant insights in their physical implications.
%
For example, an important finding of our analysis is to reveal the sensitivity of the eigenfunction fluctuations to the spectral scale as well as strength of the perturbation parameter; their spectral-dependence for the Wishart ensembles turns out to be different from the Gaussian ensembles. This may lead to crucial differences in the physical properties of the systems modeled by the two ensembles e.g  in search for their critical point, validity of ergodicity assumption, equivalence of the ensemble averaging  with their spectral averaging.
An application of our results for a specific initial condition namely Poisson statistics is discussed  in \cite{pscrit}. A detailed investigation of the results for other initial conditions will be discussed elsewhere.





Although different in detail, the approach used  here is  essentially same as the one applied in the case of a Gaussian Brownian ensemble in \cite{pswf}; the latter was based on a direct integration  of the diffusion equation for the matrix elements of the ensemble over eigenvalues.  
Due to basis-dependence of the Brownian ensemble, their eigenvalues and eigenfunctions are statistically correlated and some approximations are necessary to simplify the diffusion equations for the eigenfunction measures. In  case of a Gaussian Brownian ensembles,  the approximations used in \cite{pswf} were based on the assumption of a weak correlation between   eigenvalues and eigenfunctions  but those considered here are applicable for more generic conditions. Although our  focus in this work is Wishart Brownian ensembles, the results are presented in a form applicable also to Gaussian ensembles. This helps not only in  in a comparative study of the two ensembles  but is also helpful in extending  already available information for the Gaussian case \cite{fkpt, ap, sp} to Wishart case. 

Some of the results obtained here are used in \cite{pscrit}  for the critical point analysis of the Brownian ensembles.   But their applicability goes beyond Brownian ensembles.  Following complexity parametric formulation of the statistical fluctuations, the results  can  be extended to the system-dependent, multi-parametric  random matrix ensembles \cite{pijmp, psand, ps-all} and can help in critical point analysis of the statistics of the complex systems represented by these ensembles. The connection of Brownian ensembles with column constrained ensembles discussed in \cite{sups} also indicates the usefulness of our analysis for all those systems where the latter ensembles appear e.g  disordered systems with Goldstone symmetries, random lasers, collective spontaneous emission, Google matrix analysis etc.

\vfill\eject


\appendix



\section{Proof of equations (\ref{lmn0}-\ref{lmn2})}



The change in matrix $L=A^{\dagger}. A$ for a small increment in $Y$ can be written as 
\begin{eqnarray}
\delta L  &=&  L(Y+\delta Y) -L(Y) = A^{\dagger} (Y+\delta Y) . A(Y+\delta Y) -   A^{\dagger} (Y) . A(Y) 
\label{ltm1}
\end{eqnarray} 
Substitution of $A(Y+\delta Y) \approx A(Y) \; \left(1 -  \gamma \delta Y \right) +  \sqrt{2 \; \partial Y} \; V(Y)$ (see eq.(\ref{atm1})) in the above and keeping only terms upto first order of $\delta Y$ 
gives 
\begin{eqnarray}
\delta L
\approx  - 2 \; \gamma \;  L(Y) \; \delta Y  + 2 \;  V^{\dagger} (Y). V(Y) \; \delta Y + (A^{\dagger} V + V^{\dagger} A ) \; \sqrt{2 \delta Y}
\label{ltm2}
\end{eqnarray} 

From eq.(\ref{amn1}), the ensemble average of a matrix element, say $\langle V_{mn} \rangle=0$. Further the matrices $A$ and $V$ are assumed to be independent which implies $ \langle A^{\dagger}  . V \rangle =  \langle A^{\dagger} \rangle . \langle V \rangle =0  $. 
Taking the ensemble average of eq.(\ref{ltm2}) for a fixed $L(Y)$, the first  moment of its matrix elements $L_{mn}(Y)=\sum_{k=1}^{N_a} A_{km}^* (Y) \; A_{kn}(Y)$ can be given as 

\begin{eqnarray}
\langle \delta L_{mn} \rangle 
\approx   - 2 \; \gamma \;  L_{mn}(Y)  \; \delta Y  + 2 \;  \sum_{k=1}^{N_a}  \langle V^*_{km} (Y). V_{kn}(Y) \rangle \; \delta Y 
\label{ltm3}
\end{eqnarray}
Using eq.(\ref{amn1}), now we get
\begin{eqnarray}
\langle \delta L_{mn} \rangle 
\approx   -  2 \; \gamma \;  L_{mn}(Y)  \; \delta Y  + 2 \; \beta \; v^2\; N_a  \; \delta Y 
\label{ltm4}
\end{eqnarray}
which is same as the first moment given in eq.(\ref{lmn0}).

The 2nd moment can similarly be calculated. Using eq.(\ref{ltm2}) for a matrix element $\delta L_{mn}$ and keeping terms only upto first order of $\delta Y$, we have
\begin{eqnarray}
\delta L_{mn} \; \delta L_{kl}^*
\approx   \sum_{i,j}^{N_a} (A_{im}^* V_{in} + V_{im}^* \;  A_{in} ) \;  (A_{jk} V_{jl}^* + V_{jk} \;  A_{jl}^* ) \; \delta Y
\label{ltm5}
\end{eqnarray} 
As the correlations between $V$-matrix elements depends on $\beta$ (eq.(\ref{amn1}) gives $\langle V_{kl} \;  V_{mn} \rangle
=  v^2 \; \delta_{km} \; \delta_{ln} \; \delta_{\beta 1} $ and $\langle V_{kl} \;  V^*_{mn} \rangle
=  \beta \; v^2 \; \delta_{km} \; \delta_{ln}$ ), it is clearer to consider the cases $\beta=1$ and $2$ separately.  Taking the ensemble average after expanding the right side of eq.(\ref{ltm5})  and using relations (\ref{amn1}), we get

\noindent{\bf Case $\beta=1$} 
\begin{eqnarray}
\langle\delta L_{mn} \; \delta L_{kl}^* \rangle  &=& \langle\delta L_{mn} \; \delta L_{kl} \rangle \nonumber \\
&=& 2  \; v^2 \; \delta Y \; \sum_{i,j}^{ N} \;
\delta_{ij} \; \left[ \delta_{nl} A_{im}\; A_{jk}+ \delta_{nk} A_{im} \; A_{jl} + \delta_{ml} A_{in}\; A_{jk}+ \delta_{mk} A_{in} \; A_{jl}  \right ]  \nonumber \\
&=& 2 \; v^2 \; [L_{mk} \delta_{nl} +L_{ml} \delta_{nk} + L_{nk} \delta_{ml} + L_{nl} \delta_{mk}] \; \delta Y
\label{ltm6}
\end{eqnarray}

\noindent{\bf Case $\beta=2$} 
\begin{eqnarray}
\langle\delta L_{mn} \; \delta L_{kl}^* \rangle  
&=& 4  \; v^2 \; \delta Y \; \sum_{i,j}^{ N} \;
\delta_{ij} \; \left[ \delta_{nl} A^*_{im}\; A_{jk}+ \delta_{mk} A_{in} \; A^*_{jl}  \right ]  \nonumber \\
&=& 4 \; v^2 \; [L_{mk} \delta_{nl}  + L^*_{nl} \delta_{mk}] \; \delta Y
\label{ltm7}
\end{eqnarray}
and 
\begin{eqnarray}
\langle\delta L_{mn} \; \delta L_{kl} \rangle  
&=& 4  \; v^2 \; \delta Y \; \sum_{i,j}^{ N} \;
\delta_{ij} \; \left[ \delta_{nk} A^*_{im}\; A_{jl}+ \delta_{ml} A_{in} \; A^*_{jk}  \right ]  \nonumber \\
&=& 4 \; v^2 \; [L_{ml} \delta_{nk}  + L^*_{nk} \delta_{ml}] \; \delta Y
\label{ltm8}
\end{eqnarray}

\section{Proof of equation (\ref{enm})}

	The matrix $L$ is Hermitian in nature. 
A small change $\delta Y$ in parameter $Y$ changes $L$ and its eigenvalues $e_n$. By 
considering matrix $L+\delta L$ in the diagonal representation of matrix $L$,  
a small change $\delta e_n$ in the eigenvalue $e_n$ can be given as 
\begin{eqnarray}
\delta e_n = \delta L_{nn} +\sum_{m\not=n} {|\delta L_{mn}|^2 \over e_n-e_m}+
o((\delta L_{mn})^3)
\label{ee1}
\end{eqnarray}
where $L_{mn}=e_n \delta_{mn}$ at value $Y$ of complexity parameter. 
An ensemble averaging of the above equation  gives $\langle{\delta e_n} \rangle = \langle {\delta L_{nn}} \rangle +
\sum_{m=1,m\not=n}^{N} {\langle{|\delta L_{mn}|^2} \rangle \over e_n-e_m}$. Now using eqs.(\ref{ltm6}, \ref{ltm7}, \ref{ltm8}), this leads to 
\begin{eqnarray}
\langle{\delta e_n} \rangle 
&=& \left[ 2 \; \beta \; v^2 \; N_a -2 \; \gamma \; L_{nn} + 2 \; \beta \; v^2 \; \sum_{m=1,m\not=n}^{2N} 
 {L_{nn}+L_{mm} \over e_n-e_m}\right] \delta Y \label{eee2} \\
&=& 2 \;\beta \; v^2 \left[  N_a -{\gamma\over \beta \; v^2}  \;  e_n +   \sum_{m=1,m\not=n}^{N} 
{e_n+e_m \over e_n-e_m}\right] \delta Y
\label{ee3}
\end{eqnarray}
with $N_a$ defined above eq.(\ref{rhoh}).
Similarly  eq.(\ref{ee1}) can again be used to obtain, , upto first order of $\delta Y$,
\begin{eqnarray}
\langle{\delta e_n \; \delta e_m } \rangle=  
\langle {\delta L_{nn} \; \delta L_{mm}} \rangle =
8 \; L_{nn} \; \delta_{nm} \; \delta Y = 8  \; e_n \; \delta_{nm} \; \delta Y 
\end{eqnarray}

\section{Joint distribution of eigenvalues}


The JPDF of the eigenvalues of a Hermitian BE is discussed in detail in \cite{sp}; here we summarize only main results relevant for our analysis 
later on. 
 
A substitution of eq.(\ref{enm}) in eq.(\ref{f8}) followed by an integration over all eigenvector components leads to the diffusion equation for the joint probability distribution of the eigenvalues. 
Let $P_e(\{e_n\},Y)$ be the joint probability of finding eigenvalues $\lambda_i $  of $X$ between $e_i$ and $e_i+{\rm d}e_i$ ($i=1,2,..,N$) at $Y$, it can then be expressed as
$P_e(\{e_n\},Y)= \int \prod_{i=1}^{N}\delta(e_i-\lambda_i) 
\rho (X,Y){\rm d}X$. Using the above definition in eq.(\ref{rhoh}), the diffusion equation for the eigenvalues turns out to be
 \begin{eqnarray}
{\partial P_e\over\partial Y}   =
\sum_n {\partial \over \partial e_n}
\left[ {\partial (e_n^{\nu} \;   P_e) \over \partial e_n} + \beta
\sum_{m\not=n} {e_n^{\nu} \over  e_m-e_n} \; P_e + \beta \; a(e_n) \; P_e \right] 
\label{pev}
\end{eqnarray}
with $a(e)=\left({2\over \beta} \right)^{\nu}  \; e + a_0$, $a_0 =\nu \; (N-1-N_a)/2$  and $\nu=1$. Here for simplification, we have taken $\gamma=1$ and $v^2=1/4$. 
For comparison, it must be noted that $\nu=0$ for a Gaussian ensembles.

All spectral fluctuation measures can be derived from the set of $n$-level correlations $R_n(e_1,...,e_n;Y)$ i.e. the 
probability densities for $n$ levels  to be at $e_1,\ldots, e_n$ irrespective  of the position of 
other $N-n$ levels: 
${\it R}_n(e_1,...,e_n;Y)={N!\over (N-n)!} \; \int {\rm d}e_{n+1}...{\rm d}e_N  \; 
P(e_1,..,e_N;Y)$.   In principle, the $Y$ dependence of $R_n$ can be derived by  a direct integration of eq.(\ref{pev}). But as discussed in detail in \cite{ap,sp,fkpt}, the evolution of $R_n$ i occurs on the scales determined by $Y-Y_0 \sim \Delta_e(e)^2$ with $\Delta_e(e)$ as the local mean level spacing at energy $e$; it is therefore preferable to consider rescaled correlations $R_n(r_1,..,r_n;\Lambda)= \lim_{ N\rightarrow \infty} \;\Delta_e^n \; {\it R}_n(e_1,..,e_n; Y)$
%
%
with $r_n = {(e_n-e)\over \Delta_e(e)}$ as the rescaled spectrum. The transition in $R_n$ and therefore other spectral fluctuation measures are governed by the rescaled parameter    
\begin{eqnarray}
\Lambda_e(Y,e)={e^{\nu} \; (Y-Y_0) \over \Delta_e^2}.
\label{alm1}
\end{eqnarray}
(Appearance of $\Lambda_e$ as the transition parameter for spectral correlations can also be seen on the basis of 2nd order perturbation theory of Hermitian matrices). The $\Lambda_e$-governed evolution of unfolded correlations, from arbitrary initial condition, for both Gaussian and Wishart ensembles can be given as \cite{sp, ap, fkpt}
\begin{eqnarray}
{\partial R_n \over\partial \Lambda_e} &=& \sum_j {\partial^2 R_n\over \partial r_j^2}-\beta \sum_{j\not=k} {\partial \over \partial r_j} \left({R_n \over {r_j-r_k}}\right) -\beta \sum_j {\partial \over \partial r_j} \int_{}^{} {R_{n+1} \over {r_j-r}} \; {\rm d}r. 
\label{rn}
\end{eqnarray}
%
It is worth emphasizing here that  eq.(\ref{rn}) is based on the assumption that 
the correlations are  localized around spectral scale $e$ (i.e the local stationarity condition $\sum {\partial R_n \over \partial r_j}=0$ alongwith the assumption $R_1(e_k) \approx R_1(e)$ for all $k=1,\ldots, n$)  and are separable: $R_{n+1}(r_1, \ldots, r_n, r_{n+1}) \to R_n(r_1, \ldots, r_n)$ as $r_{n+1} \to \infty$ \cite{fkpt}.  Here the limits $\Lambda_e=0, \infty$ correspond to the initial and the stationary state for the BE, respectively.   The stationary solution can be obtained by substituting $  {\partial R_n \over \partial \Lambda_e}=0$ in eq.(\ref{rn}) which leads to 
$ {\partial R_n\over \partial r_j}-\beta \sum_{k; k\not=j}  \left({R_n \over {r_j-r_k}}\right) -\beta  \int_{}^{} {R_{n+1} \over {r_j-r}} \; {\rm d}r = 0.$ 
%
For later reference, it is worth noting that, for small $r_j$ values, the integral term in eq.(\ref{rn}) can be neglected but it is crucial to  obtain the expected limiting behaviour $R_2(r_1, r_2) \to 1$ for $|r_1-r_2| \to \infty$ \cite{fkpt}.  

%


The solution of eq.(\ref{rn}) for any non-zero, finite $\Lambda_e$ corresponds to an intermediate, non-equilibrium  statistics. 
As the energy-dependence of  $\Lambda_e$ originates from the level-density $R_1$, it is necessary to consider the parametric-variation
of $R_1$ too. Ignoring the 2nd derivative (being of $O(1/N)$ with respect to other terms), the $Y$-dependent evolution 
of $R_1$ reduces to Dyson-Pastur equation \cite{fkpt,ap,sp,apps}
\begin{eqnarray}
{\partial R_1 \over\partial Y} &=& \left({\beta\over 2}\right)^{1-\nu}  {\partial \over \partial e} \left( e^{\nu} \; R_1(e, Y) \right) - {\beta\over 2} \;  {\partial \over \partial e} \int_{spr} { e^{\nu} \; R_1(e) R_1(e') \over {e-e'}} \; {\rm d}e' 
\label{r1}
\end{eqnarray}
with subscript ${\it spr}$ implying the spectral region, extending from $-\infty  \rightarrow \infty$ in 
the Gaussian case and $0 \rightarrow \infty$ for Laguerre case. An important point clearly indicated by the above equation is that $R_1(e)$ is non-stationary as well as non-ergodic; as discussed in \cite{pscrit}, this plays a crucial role in defining the criteria for criticality of the spectral statistics.

As discussed in \cite{sp}, eq.(\ref{r1}) can be solved by defining the resolvant $G(z;\tau)=\int_R  \frac{R_1(x;\tau)}{z-x} \; {\rm d}x$ which satisfies $G(x+i0;\tau)=\int_R \frac{R_1(y;\tau)}{x-y} \; {\rm d}y - i\pi R_1(x;\tau)$. For Gaussian case, the solution of eq.(\ref{r1}) for many initial conditions is already known \cite{sp, shapiro}. For Wishart case, the solution can be obtained by noting the following relation: 
the diffusion equation for the resolvent $G$ for Gaussian and Wishart cases are given by  eq.(38) and eq.(39) of \cite{sp}, respectively; the latter  can be reduced to the former  by replacing $z \rightarrow \frac{\beta^2}{2} z^2$ and $\tau \rightarrow 2 \tau$. By applying the same transformation, therefore, $R_1(z)$ for Wishart BE can be obtained from the known $R_1(z)$ results for Gaussian BE.

\section{Derivation of eq.(\ref{qaba2})}

Consider the integral 
\begin{eqnarray}
Q_{mn;k}^{r s}  &=&  \left({1\over 2}\right)^{\nu} \; \sum_{j=1; j \not=k}^N\int {(e_k+e_j)^{\nu} \over (e_k-e_j)^s} \; (z_{mj} z^*_{nj})^r \; {P_{N2}} \; {\rm d}e_j {\rm D}^{\beta} Z_j, 
\label{qi1}
\end{eqnarray}

The correlation between eigenvalues in random matrix ensembles are known to decay rapidly with their separation, with those beyond a few mean level spacings are uncorrelated. 
For distances $|e_k-e_j| > N_k \Delta_e$, with $N_k \sim O(1)$,  one can then approximate  ${P_{N2}} (Z_k, Z_j, e_k, e_j)  \approx  P_{N1}(Z_k,e_k) \; P_{N1}(Z_j, e_j) $. The above integral can now be rewritten as 
 \begin{eqnarray}
Q_{mn;k}^{r s}  &=& \left({1\over 2}\right)^{\nu} \;  \sum_{j=1; j\not=k}^N (E_1+ E_2 + E_3 )  \label{q1}\\ 
E_{1} &=& \int_{-\infty}^{e_k-\Omega_k/2} 
 {(e_k+e_j)^{\nu} \over (e_k-e_j)^s} \; (z_{mj} z^*_{nj})^r \; P_{N1}(Z_k,e_k) \; P_{N1}(Z_j, e_j) \; {\rm d}e_j {\rm D}^{\beta} Z_j, \\
E_{2} &=& \int_{e_k-\Omega_k/2}^{e_k+\Omega_k/2} 
 {(e_k+e_j)^{\nu} \over (e_k-e_j)^s} \; (z_{mj} z^*_{nj})^r \; {P_{N2}} (Z_k, Z_j, e_k, e_j) \; {\rm d}e_j {\rm D}^{\beta} Z_j, \label{e2}\\
E_{3} &=& \int_{e_k+\Omega_k/2}^{\infty} 
 {(e_k+e_j)^{\nu} \over (e_k-e_j)^s} \; (z_{mj} z^*_{nj})^r \; P_{N1}(Z_k,e_k) \; P_{N1}(Z_j, e_j) \; {\rm d}e_j {\rm D}^{\beta} Z_j, 
\label{qi2}
\end{eqnarray}
where $\Omega_k$ is a spectral range of the order of few mean level-spacings: $\Omega_k=N_k \; \Delta_k$ with $\Delta_k(e_k)$ as the local mean level spacing and $N_k$ as the number of eigenvalues in this range. 

Using the definition  $\langle  (z_{mj} z^*_{nj})^r  \;  \rangle = {N\over R_1(e_j) } \; \int 
 (z_{mj} z^*_{nj})^r  \; P_{N1}(Z_j, e_j) \; {\rm D}^{\beta} Z_j$ with $R_1(e) $ as the ensemble averaged level density i.e $R_1(e)= N \int P_{N1}(Z,e) \; {\rm D}^{\beta}Z$ with $\int_{-\infty}^{\infty} R_1(e) \; {\rm d}e =N$, the integrals  $E_1, E_3$ can further be written as 
 \begin{eqnarray}
E_{1} &=& {(-1)^s\over N} \;  \langle  (z_{mj} z^*_{nj})^r  \;  \rangle \; \; P_{N1}(Z_k,e_k) \int_{-\infty}^{-\Omega_k/2} 
 {(2 e_k+y)^{\nu} \over y^s} \;\; R_1(e_k+y) \; {\rm d}y, \\
E_{3} &=& {1 \over N} \; \langle  (z_{mj} z^*_{nj})^r \rangle \; P_{N1}(Z_k,e_k) \int_{\Omega_k/2}^{\infty} 
 {(2 e_k+y)^{\nu} \over y^s} \; R_1(e_k+y) \; {\rm d}y, 
\label{qi3}
\end{eqnarray}

Due to confinement of the eigenvalues, $R_1$ decays for large spectral-ranges and main contribution to the integral in $E_1$ 
comes from the neighborhood of  $y \sim -\Omega_k/2$.  One can then approximate $E_1$ as
\begin{eqnarray}
E_{1} &=&  {1\over N} \; \langle  (z_{mj} z^*_{nj})^r  \;  \rangle \; \; P_{N1}(Z_k,e_k) \;
 {2^s \; (2 e_k-\Omega_k/2)^{\nu} \over \Omega_k^s} \;\int_{-\infty}^{-\Omega_k/2}  \; R_1(e_k+y) \; {\rm d}y, 
\label{qi4}
\end{eqnarray}
Similarly $E_3$ becomes
\begin{eqnarray}
E_{3} &=& {1\over N} \; \langle  (z_{mj} z^*_{nj})^r  \;  \rangle \; \; P_{N1}(Z_k,e_k) \;
 {2^s \; (2 e_k+\Omega_k/2)^{\nu} \over \Omega_k^s} \;\int_{\Omega_k/2}^{\infty}  \; R_1(e_k+y) \; {\rm d}y, 
\label{qi5}
\end{eqnarray}
Now as $\int_{-\infty}^{-\Omega_k/2}  \; R_1(e_k+y) \; {\rm d}y +\int_{\Omega_k/2}^{\infty}  \; R_1(e_k+y) \; {\rm d}y = 
\int_{-\infty}^{\infty} R_1(e_k+y) \; {\rm d}y -\int_{-\Omega/2}^{\Omega/2} R_1(e_k+y) \; {\rm d}y \approx N-N_k$, 
we have 
 \begin{eqnarray}
E_{1}+E_{3} \approx    {2^s \; (2 e_k)^{\nu} \over \Omega_k^2}  \; \langle  (z_{mj} z^*_{nj})^r  \;  \rangle \; \; P_{N1}(Z_k,e_k) \;
\label{qi6}
\end{eqnarray}
(Here the term $(2 e_k+\Omega_k/2)^{\nu}$ is approximated as $(2 e_k)^{\nu} $ which is valid only spectral ranges $e_k \gg \Omega_k$ and therefore for  the bulk spectrum of Wishart ensemble or away from the hard spectrum edge $e=0$. For Gaussian case with $\nu=0$, this approximation is not needed). 


To calculate $E_{2}$, we note that the integral over $y$  in eq.(\ref{ee2})  is confined over a very small spectral range  $\Omega_k$ around $e_k$. 
As the average correlation between components of an eigenfunction is expected to be almost same as another eigenfunction if their eigenvalues are approximately equal. Thus for $e_j \in \Omega_k$, one can approximate
$\overline{\langle (z_{nj} z_{mj}^*\rangle}_{e_k} \approx \overline{\langle z_{nk} z_{mk}^*\rangle}$ 
where $\overline{\langle z_{nj} z_{mj}^* \rangle}_{e_k}$ is the ensemble as well as spectral averaged local correlation of an eigenstate with its energy close to $e_k$: $\overline{\langle z_{nj} z_{mj}^* \rangle}_{e_k} = {1\over \Omega_k} \; \int_{\Omega_k}  \langle z_{nj} z_{mj}^* \rangle \; {\rm d}e$. This leads to
\begin{eqnarray}
E_{2} \approx (-1)^s \; \overline{\langle (z_{nk} z_{mk}^*)^r \rangle} \; \int_{-\Omega_k/2}^{\Omega_k/2} 
 {(2 e_k+y)^{\nu} \over y^s} \; {\mathcal P}_{N2} (Z_k, e_k, y) \; {\rm d}y  \label{ee2} 
\end{eqnarray}
where ${\mathcal P}_{N2} (Z_k, e_k, y) = \int  \; P_{N2} (Z_k, Z_j, e_k, y) \; {\rm D}^{\beta} Z_j$.  
Now expanding ${\mathcal P}_{N2}(Z_k, e_k, e_k+y)$ in Taylor's series around $y =0$, eq.(\ref{ee2}) can be approximated as
\begin{eqnarray}
E_{2} &\approx & (-1)^ s \; \overline{\langle (z_{nk} z_{mk}^*)^r \rangle} \; \sum_{n=0}^{\infty}  {\alpha_n\over n!}  \;
 {{\rm d}^n{\mathcal P_{N2}} \over {\rm d}y^n}\mid_{y=0} 
\label{ef2} 
\end{eqnarray}
where $\alpha_n =\int_{-\Omega_k/2}^{\Omega_k/2}  (2 e_k+y)^{\nu} \; y^{n-s} \; {\rm d}y$. 
Neglecting  terms with higher powers of  $\Omega_k$,  the above leads to, for $s=2$,  
$E_{2} \approx  \alpha_0 \;  {\mathcal P}_{N2}(Z_k, e_k, 0) = -{4 \; (2 e_k)^{\nu} \over \Omega_k}  \;  {\mathcal P}_{N2}(Z_k, e_k, 0) $. 
Similarly, for $s=1$, $E_{2} \approx  \left(2 \nu e_k + {{\rm d}{\mathcal P_{N2}} \over {\rm d}y}\mid_{y=0} \right)\; \Omega_k  \;  {\mathcal P}_{N2}(Z_k, e_k, 0) $. 

Following the definition $P_{N1}(Z_k, e_k) = \int_{-\infty}^{\infty}  {\mathcal P}_{N2}(Z_k, e_k, e_k+y) \; {\rm d}y $, one can write ${\mathcal P}_{N2}(Z_k, e_k, 0)  \propto P_{N1}(Z_k, e_k) $. As $\Omega_k \sim E_c \ll 1$, the contribution from $E_{2}$ is negligible as compared to $E_{1}, E_{3}$. 
%
 %
%
%
 Substitution of eq.(\ref{qi6}) in eq.(\ref{q1}) now leads to
\begin{eqnarray}
Q_{mn;k}^{r s} &\approx &   {\mathcal K}_s \; \; \left(\overline{\langle z_{nk} z_{mk}^* r\rangle} \right)^r  \; P_{N1}(Z_k, e_k)
\label{qabaa2}
\end{eqnarray}
where 
\begin{eqnarray}
{\mathcal K}_s(e_k) &=&   \left({2\over \Omega_k}\right)^s \; N \;  e_k^{\nu}, \label{kka} 
\end{eqnarray}

 
\end{document}